\documentclass[twocolumn]{aa}
\usepackage{graphicx,epsf}
\begin{document}
\renewcommand{\topfraction}{0.85}
\renewcommand{\bottomfraction}{0.7}
\renewcommand{\textfraction}{0.15}
\renewcommand{\floatpagefraction}{0.90}
   \title{H.E.S.S. Observations of PKS 2155$-$304}
\author{F. Aharonian\inst{1}
 \and A.G.~Akhperjanian \inst{2}
 \and K.-M.~Aye \inst{3}
 \and A.R.~Bazer-Bachi \inst{4}
 \and M.~Beilicke \inst{5}
 \and W.~Benbow \inst{1}
 \and D.~Berge \inst{1}
 \and P.~Berghaus \inst{6} \thanks{Universit\'e Libre de
 Bruxelles, Facult\'e des Sciences, Belgium}
 \and K.~Bernl\"ohr \inst{1,7}
% ...
}
 
\institute{
Max-Planck-Institut f\"ur Kernphysik, Heidelberg, Germany
\and
 Yerevan Physics Institute, Armenia
\and
University of Durham, Department of Physics, U.K.
\and
Centre d'Etude Spatiale des Rayonnements, CNRS/UPS, Toulouse, France
\and
Universit\"at Hamburg, Institut f\"ur Experimentalphysik, Germany
\and
Physique Corpusculaire et Cosmologie, IN2P3/CNRS, Coll{\`e}ge de France, Paris, France
\and
Institut f\"ur Physik, Humboldt-Universit\"at zu Berlin, Germany
% ...
}
 
   \offprints{W. Benbow}

   \date{Received August 17, 2004; Accepted September 27, 2004}

   \abstract{
The high-frequency peaked BL Lac 
PKS 2155$-$304 at redshift $z$=$0.117$ has been detected with high 
significance ($\sim$45$\sigma$) at energies greater than
160 GeV, using the H.E.S.S. stereoscopic array of imaging 
air-Cherenkov telescopes in Namibia. A strong signal is found in 
each of the data sets corresponding to the dark periods of July 
and October, 2002, and June-September, 2003.   
The observed flux of VHE gamma rays shows variability on time scales of months, days, 
and hours.  The monthly-averaged integral flux above 300 GeV varies 
between 10\% and 60\% of the flux observed from the Crab Nebula.  Energy spectra are measured for
these individual periods of data taking and are characterized by a steep power law with a 
time-averaged photon index of $\Gamma=3.32\pm0.06$.  
An improved $\chi^2$ per degree of freedom
is found when either a power law with an exponential cutoff energy
or a broken power law are fit to the time-averaged energy spectrum.
However, the significance of the improvement is marginal ($\sim$2$\sigma$).  
The suggested presence of features in 
the energy spectrum may be intrinsic to the emission from the blazar, 
or an indication of absorption of TeV gamma rays by the extragalactic infrared background light.

   }

   \maketitle
%

\section{Introduction}
The high-frequency peaked BL Lac (HBL) object PKS 2155$-$304 ($z=0.117$) was discovered by 
the {\it HEAO 1} X-ray satellite (\cite{heao1};~\cite{heao2}) at a position consistent with a 
poorly localized detection by the {\it Ariel V} satellite ~\cite{arielV}.  PKS 2155$-$304 is now known to
be one of the brightest extragalactic X-ray sources in the sky and as a result other X-ray satellites, 
including {\it ROSAT}, {\it BeppoSAX}, {\it RXTE}, and {\it Chandra} 
(see, e.g., ~\cite{ROSAT_det}; ~\cite{Beppo_det}; ~\cite{RXTE_det}; ~\cite{Chandra_det}), have detected it 
on a regular basis. PKS 2155$-$304 is well studied and has a history of strong broad-band variability. 
It is associated with a compact, flat-spectrum radio source, and exhibits an essentially 
featureless continuum from radio to X-ray frequencies.  The maximum power emitted by PKS 2155$-$304 is between 
the UV and soft X-ray range, and it is the brightest BL Lac detected in the UV regime ~\cite{history1}.  
Gamma-ray emission in the energy range 30 MeV to 10 GeV was 
detected from PKS 2155$-$304 by the {\it EGRET} detector 
aboard the {\it Compton Gamma Ray Observatory} satellite ~\cite{egret}. The {\it EGRET} observations indicated a hard energy spectrum 
with integral power-law photon index of $1.71\pm0.24$.  All of these characteristics make PKS 2155$-$304 a 
likely emitter of $>$100 GeV gamma rays.  In fact, it was detected at energies greater than 300 GeV in 1996 and 1997 by the 
{\it University of Durham Mark 6 Telescope} as reported in ~\cite{durham1}.  However, the object was not detected by the same 
instrument in 1998 ~\cite{durham2}.  Furthermore, PKS 2155$-$304 was not detected by the {\it CANGAROO} experiment during
observations made in 1997 ~\cite{cangaroo1}, 1999 ~\cite{cangaroo2}, 2000 or 2001 ~\cite{cangaroo3}.  The 
reported upper limit from {\it CANGAROO} in 1997 is consistent with the flux detected by Durham, and
the lack of a detection of PKS 2155$-$304 at TeV energies in subsequent years is not inconsistent 
with the fact that such emission from AGN is known to be highly variable.

The present confirmation of the detection of VHE gamma-rays from PKS 2155$-$304 by the High Energy Stereoscopic System (H.E.S.S.)
along with observations at other wavelengths will yield considerable insight into the mechanisms 
for TeV gamma-ray emission from blazars, help delineate the differences between low-frequency peaked BL Lacs and HBLs, 
and will assist in the understanding of the extragalactic infrared background light (EBL).  
This latter point is strengthened by the fact that PKS 2155$-$304 is the second most distant object 
detected at TeV energies and therefore its energy spectrum may exhibit characteristics, such as steepening of the spectrum 
and a cutoff, caused by absorption of VHE gamma rays via pair production on the EBL.  The most distant object
detected at TeV energies, 1ES 1426$+$428 ($z=0.129$), showed features in its energy spectrum which have been interpreted 
as consequences of such absorption ~\cite{hegra_1426a}.  In addition, variability studies using detailed light curves
obtained from observations of PKS 2155$-$304 can place strong constraints on the physical modeling of blazars.

\section{H.E.S.S. Detector}
The H.E.S.S. experiment has been operating since June, 2002, in the Khomas Highlands of Namibia
(23$^{\circ}$ 16' 18'' S, 16$^{\circ}$ 30' 1'' E, 1835 m above sea level).  The
detector consists of a system of four imaging air-Cherenkov telescopes in a square of 120 m side.  Each individual
telescope is an alt-az mount Davies-Cotton reflector ~\cite{dav_cot_ref} with a flat-to-flat width of 13 m,
and has a camera mounted at the focal length of 15 m.  The total mirror area of each telescope is 107 m$^{2}$, segmented 
into 382 individual round (60 cm diameter) front-aluminized glass mirrors.  The H.E.S.S. cameras
provide a 5$^{\circ}$ field of view and contain 960 individual photomultiplier (PMT) pixels subtending
0.16$^{\circ}$ each, with Winston cone light concentrators.  The camera is modular in design,
housing 60 drawers of 16 PMTs each, and contains all the necessary electronics for operation, triggering, and readout.  The trigger
electronics divide the camera into overlapping 64 PMT sectors (4 adjacent drawers) with a
trigger requirement that a sector has a minimum number
of pixels with a signal above a threshold in photoelectrons (PEs) coincident in an effective 
$\sim$1.3 ns trigger window.  Upon receiving a camera
trigger, the signal stored in analog memories from each of the PMTs, sampled at 1 GHz, is integrated within a 16 ns window.  
Once a camera has triggered, a signal is sent out via an optical fiber to a central trigger system ~\cite{cent_trig} which 
allows for a multiple telescope coincidence requirement. More details on H.E.S.S. can be found in ~\cite{HESS1}; ~\cite{HESS2}; ~\cite{HESS3}.

\section{Observations}

   \begin{table*}
      \caption{Shown are the dark periods in which PKS 2155$-$304 was observed,
	the configuration of H.E.S.S. during those observations, the telescope
	multiplicity required to trigger the array, the number of telescopes available,
	the sector trigger requirement (number of pixels above a threshold in PEs),
	the observation mode, the number of on-source runs,
	the dead time corrected observation time, the system rate, the dead time percentage,
	and the mean zenith angle of the observations ($Z_{obs}$).}
         \label{obstime}
	\centering
         \begin{tabular}{c   c   c   c   c   c   c   c   c   c}
            \hline\hline
            \noalign{\smallskip}
	    Dark & & N$_{\mathrm{tel}}$ & Sector & Obs. & No. & Obs. Time & Syst. Rate & Dead  Time & Z$_{\mathrm{obs}}$\\
            Period & Configuration & Mult.,Tot. & Trigger & Mode & Runs & [hrs  live] & [Hz] & \% & [$^{\circ}$] \\
            \noalign{\smallskip}
            \hline
            \noalign{\smallskip}
            07/2002 & Mono (30 Drawers) & 1,1 & 4 $>$ 6.7 & On $-$ Off & 8 &  2.5 & 130 & 24 & 15\\
            10/2002 & Mono (52 Drawers) & 1,1 & 4 $>$ 6.7 & On $-$ Off & 16 & 4.3 & 230 & 37 & 16\\
            11/2002 & Mono & 1,1 & 4 $>$ 6.7 & On $-$ Off & 4 & 0.8  & 250 & 53 & 35\\
            06/2003 & Software Stereo & 2,2 & 4 $>$ 6.7 & Wobble & 32 & 10.8  & 60 & 23 & 13\\
            07/2003 & Hardware Stereo & 2,2 & 3 $>$ 5.3 & Wobble & 55 & 22.1  & 100 & 7 & 24\\
            08/2003 & Hardware Stereo & 2,2 & 3 $>$ 5.3 & Wobble & 45 & 19.6  & 100 & 7 & 14\\
 	    09/2003 & Hardware Stereo & 2,2 & 3 $>$ 5.3 &  Wobble & 2 & 0.9 & 120 & 8 & 8\\
	    09/2003 & Hardware Stereo & 2,3 & 3 $>$ 5.3 &  Wobble & 6 & 2.1 & 170 & 12 & 18\\
            \noalign{\smallskip}
            \hline
       \end{tabular}
   \end{table*}

The H.E.S.S. observations of PKS 2155$-$304 were made while the system was under construction, 
therefore the data cover various telescope configurations and
trigger criteria.  Observations in 2002 were made with just one telescope.  In January 2003, a second telescope was added to the
array and observations in 2003 consist mostly of a two-telescope configuration.  The exception to this is in September 2003, 
when a third telescope was added to the array.  The fourth and final telescope was 
added to the array in December 2003, after the end of the observation season for PKS 2155$-$304.

The trigger for H.E.S.S. has similarly evolved.  During the observations in July, October, November 2002,
and June 2003, the sector trigger consisted of a requirement for 4 pixels each with more than 6.7 PEs.  However, the 
July and October 2002 observations were made with only the inner 30 and 52 drawers, 
respectively, participating in the trigger while the entire camera was used for image collection.
For all later observations the cameras were used fully.  The system level trigger was installed in July 2003.
Before this time two-telescope data were taken with each telescope separately, 
and the stereo multiplicity requirement (two telescopes) was performed off-line (``Software Stereo'')
using GPS time stamps.  
After the installation of the central trigger system the stereo multiplicity requirement 
was performed in the hardware (``Hardware Stereo'').  As the system rate was considerably lower 
due to a two-telescope multiplicity requirement, 
the sector trigger requirement could be lowered to 3 pixels with more than 5.3 PEs, allowing for a lower energy 
threshold.  This increased the individual telescope rates, from ~$\sim$250 Hz to ~$\sim$800 Hz, while maintaining
a reduced system dead time. 

Table~\ref{obstime} gives details
of the observations of PKS 2155$-$304 by H.E.S.S. which pass conservative run selection criteria.  These 
run selection criteria remove runs for which the sky was not clear and where the 
telescopes were not operating within specified requirements (e.g., trigger rate stability, more than 95\% of the
camera pixels operational).  All the single telescope observation runs were taken in {\it On-Off} mode.  
Here, data were taken on source for 25 minutes, preceded or followed
by off-source observations, used to estimate the background.  These off-source observations differ by $\pm$30 minutes 
in right ascension (RA) from the actual source position.  Given an approximately 5 minute (reduced to 2 minutes after 2002) 
transition time for the data acquisition system, 
this ensures that the on and off runs have identical azimuth and altitude profiles.
The stereo data were taken in 28 minute runs using {\it Wobble} mode.  In this mode,
the source direction is positioned $\pm$0.5$^{\circ}$ in declination relative to the center of the field of view of the camera
during observations.  The sign of the offset is alternated in successive scans to reduce systematic effects.
Due to the large field of view of the H.E.S.S. cameras, use of {\it Wobble} mode 
allows for both on-source observations and simultaneous estimation 
of the background induced by charged cosmic rays, since the background can be estimated
from different regions in the same field of view.  This eliminates the need for off-source observations and therefore
doubles the amount of time available for on-source observations.

\section{Technique}

   \begin{table*}
      \caption{The selection cuts applied to the data and the percentage of $\gamma$-ray and background events retained by those cuts.}
         \label{thecuts}
	 \centering
         \begin{tabular}{l   c   c   c   c   c   c   c   c   c   c   c   c}
            \hline\hline
            \noalign{\smallskip}
            & & MRSL & MRSL & MRSW & MRSW & C.O.G. & Size & Length/Size & N$_{\mathrm{tel}}$ & $\theta^2$ & $\gamma$ & BG\\
	    Configuration &  N$_{\mathrm{tel}}$ & min & max & min & max & max & min & max & min & max & \% & \% \\
	    & & [$\sigma$] & [$\sigma$] & [$\sigma$] & [$\sigma$] & [mrad] & [PE] & [mrad/PE] & & [deg$^{2}$] & & \\
            \noalign{\smallskip}
            \hline
            \noalign{\smallskip}
            Mono & 1 & $-5.0$ & 3.4 & $-4.2$ & 1.0 & 35 & None & 0.016 & 1 & 0.035 & 20 & 0.012\\
	    Stereo(Soft.) & 2 & $-2.2$ & 3.2 & $-10.0$ & 1.6 & 35 & 75 & None & 2 & 0.030 & 72 & 0.036\\
            Stereo(Hard.) & 2 & $-1.8$ & 2.0 & $-10.0$ & 1.0 & 35 & 55 & None & 2 & 0.025 & 48 & 0.029 \\
            Stereo(Hard.) & 3 & $-1.8$ & 2.0 & $-10.0$ & 1.0 & 35 & 55 & None & 2 & 0.025 & 54 & 0.041 \\
            \noalign{\smallskip}
            \hline
         \end{tabular}
   \end{table*}
  
\subsection{Event Reconstruction}	
The analysis of the data passing the run selection criteria proceeds in the following steps:
First the images are calibrated ~\cite{calib_paper} and then ``cleaned'' to remove noise from the image.  
The image cleaning is done using a two-stage tail-cut procedure which requires a 
pixel to have a signal greater than 10 PEs and a neighboring pixel to have a signal larger than
5 PEs. Also pixels greater than 5 PEs are included if they have a neighbor greater than 10 PEs.
After this image cleaning is performed, 
the moments of the shower image are parameterized using a Hillas-type analysis ~\cite{hillas}.  
The shower geometry is reconstructed using stereoscopic techniques ~\cite{stereo_tech} with a typical angular resolution
of $\sim$0.1$^{\circ}$ and an average accuracy of $\sim$10 m in the determination of the shower core location. 
To ensure that the analyzed images are not truncated by the edge of the camera, only images which pass 
a distance cut requiring the image center of gravity to be less than 2$^{\circ}$ from the center 
of the camera are used in the reconstruction.  
For the stereo configuration a minimum of 2 telescopes meeting 
this criterion are required for a successful reconstruction.  In addition, at least two telescopes are each required
to exceed a minimum total signal to ensure that the images are well reconstructed.
The analysis techniques are similar for the mono-configurations, although a
necessary difference arises in the determination of the shower geometry.
In the mono-telescope case, the impact parameter and direction of the shower axis are estimated directly from the image. 
To do this, the angular distance, along the major axis of the reconstructed ellipse, from the image center of gravity 
to the point of origin of the event is estimated using the image parameter construction length/log(size) ~\cite{mono_geom}.
                                                                                   
\subsection{Background Rejection}
                                   
After the event reconstruction, the much more numerous cosmic-ray background events
are rejected using cuts on mean reduced scaled width (MRSW)
and length (MRSL) parameters.  These parameters are defined as the
mean of the difference in standard deviations for each telescope of the width (length) observed in the image 
from that which is expected from gamma-ray simulations ($<$width$>$ and $\sigma$) based on image intensity,
reconstructed impact parameter and zenith angle of observations.  The equation for MRSW is as follows:
\begin{equation}
\label{scale_pars}
\begin{array}{c}
MRSW =	\frac{1}{N_{\mathrm{tel}}} \hspace{0.5ex} 
	\sum_{i=0}^{N_{\mathrm{tel}}} \frac{\mathrm{width}_{i}  \hspace{0.5ex} -  \hspace{0.5ex} <\mathrm{width}>_{i}}{\sigma_{i}} \hspace{0.5ex} , 
\end{array}
\end{equation} 
\noindent 
and is similar for MRSL.
Images from muons are rejected by requiring events be coincident in at least 2 telescopes in the stereo configuration 
or by using a cut on length divided by size of the image, after image cleaning, for the mono-configuration ~\cite{lovers}.  
A cut on $\theta^{2}$, the square of the angular difference between the reconstructed shower position and
the source position, is applied and is equivalent to placing the data into a round bin centered on the source position. 
All the cuts (shown in Table~\ref{thecuts}) are optimized {\it a priori} (simultaneously) to yield the maximum
expected significance per hour of observation.  The data used for optimization consists of Monte Carlo gamma-ray simulations 
at a zenith angle of 20$^{\circ}$ with a Crab-like 
energy spectrum (spectral index=$2.6$, 1/2 Crab flux) and real off-source data.  The percentage of gamma-rays and
cosmic rays retained by the cuts is also shown in Table~\ref{thecuts}.   The significance expected (and observed)
is not strongly dependent on the exact values of the cuts used. 
For each of the various hardware configurations a different set of cuts is used, all optimized on Monte Carlo simulations 
and comparable off-source data sets. 

\subsection{Background Estimation}
For mono-telescope observations the background is estimated from the number of events passing the selection cuts at the center
of the field of view during off-source observations.  The background is then normalized using the ratio
of dead time corrected observation time.  For the stereo ({\it Wobble}) observations 
the background is estimated using all events passing cuts in a ring around the source location.  
The distance from the source position to the radial center of the ring is 0.5$^{\circ}$, and the width of the ring is 
adjusted such that the area of the ring is approximately seven times the area of the on-source region. The normalization of the
off-source event total is corrected for the radial acceptance of the camera. The use of a
larger background region in this ring technique reduces the relative statistical error on the background measurement.  
After the number of on-source and background events passing the selection cuts are determined, 
the significance of the excess is calculated following the method of Equation (17) in \cite{lima}.  

\section{Detection of PKS 2155$-$304}

\begin{figure}
  \begin{center}
  \epsfxsize=8.7cm
  \epsffile{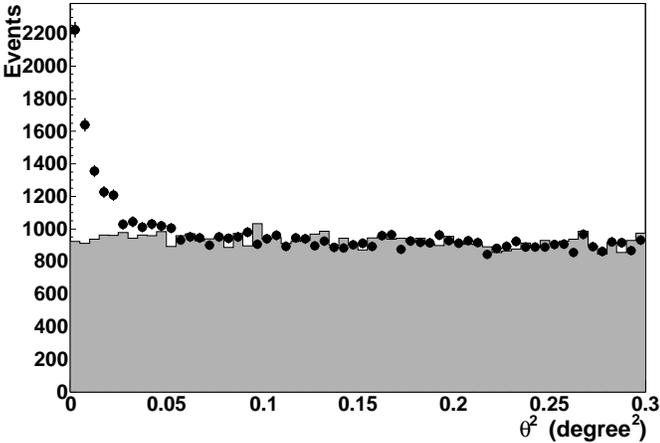}
  \end{center}
  \caption{The distribution of $\theta^{2}$ for 
	on-source events (points) and normalized 
	off-source events (shaded) from observations of PKS 2155$-$304
	in the two-telescope hardware stereo configuration. 
	For this plot, the background is taken from a region of
	comparable area located at the opposite offset position in the field of view.  
	This causes the significance shown to be reduced due to the smaller 
	area used for the background estimation.}
  \label{h2_thtsq}
  \end{figure}

\begin{figure}
  \begin{center}
  \epsfxsize=8.7cm
  \epsffile{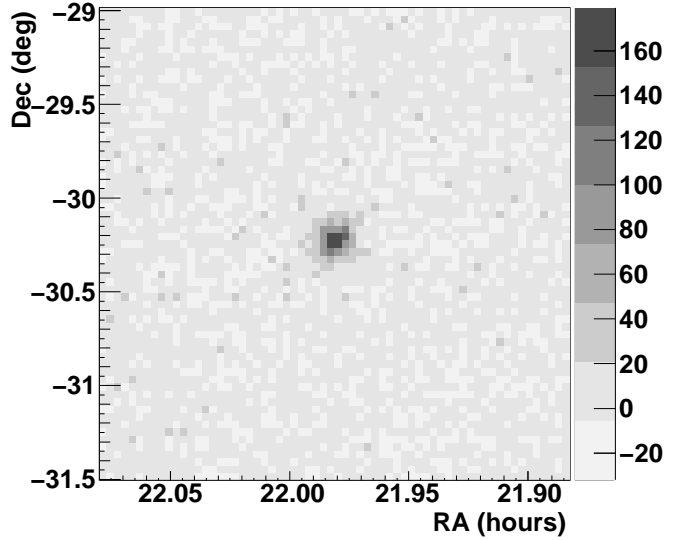}
  \end{center}
  \caption{A two dimensional distribution of excess events observed 
	in the direction of PKS 2155$-$304
	for the two-telescope hardware stereo configuration.}
  \label{h2_2D}
  \end{figure}

PKS 2155$-$304 is strongly detected in all the dark periods in which it was observed with the
exception of November 2002 where the exposure was less than one hour.
Table~\ref{results} shows the results of the H.E.S.S. observations for each of the individual
dark periods. The total significance of the excess for all observations
is 44.9$\sigma$. Figure~\ref{h2_thtsq} shows the on-source and normalized off-source 
distributions of $\theta^{2}$ for all observations in the two-telescope hardware stereo 
configuration\footnote{This configuration represents the major share of the data; the shapes of the
respective curves are similar for the other configurations.}.  The background is 
flat in $\theta^{2}$ as expected, and there is a clear excess at small values of 
$\theta^{2}$ corresponding to the observed signal. Figure~\ref{h2_2D} shows a two-dimensional sky map of the excess observed in
the direction of PKS 2155$-$304 for the same configuration.  The bins are not correlated and represent the actual distribution
of observed gamma-rays (and some background events) on the sky.  
A fit of the peak in Figure~\ref{h2_2D} to a two-dimensional Gaussian 
finds the shape of the excess to be characteristic of a point source.  The fit peak is located  
($\Delta_{RA}=9\pm12_{\mathrm{stat}}$ arcsec, $\Delta_{dec}=14\pm10_{\mathrm{stat}}$ arcsec)
from the position of PKS 2155$-$304, consistent with the position of
the blazar as expected (systematic errors on the pointing are less than 20 arcsec in RA and declination ~\cite{pointing}).  

  \begin{table}
      \caption{Shown are the number of on-source and off-source events passing the cuts,
	the normalization for the off-source events, the observed excess from PKS 2155$-$304, 
	and the significance of the excess for each of the dark periods.}
         \label{results}
	 \centering
         \begin{tabular}{c c c c c c c}
            \hline\hline
            \noalign{\smallskip}
	    Dark & & & & & & Sig.\\
	    Period & N$_{\mathrm{tels}}$ & On & Off & Norm & Excess & [$\sigma$]\\
            \noalign{\smallskip}
            \hline
            \noalign{\smallskip}
            07/2002 & 1 & 637 & 234 & 1.063 & 388 & 13.0 \\
            10/2002 & 1 & 865 & 519 & 1.065 & 312 & 8.2 \\
            11/2002 & 1 & 80 & 90 & 0.896 & $-0.65$ & $-0.1$ \\
	    06/2003 & 2 & 1396 & 4619 & 0.152 & 694 & 21.1 \\
            07/2003 & 2 & 3169 & 13643 & 0.147 & 1164 & 22.1 \\
            08/2003 & 2 & 3369 & 12955 & 0.147 & 1459 & 27.7 \\
 	    09/2003 & 2 & 185 & 697 & 0.148 & 82 & 6.7 \\
 	    09/2003 & 3 & 1005 & 3946 & 0.147 & 425 & 14.7 \\
	    \hline
	    Total & Stereo & & & & & 43.8 \\
	    Total & All    & & & & & 44.9 \\
            \noalign{\smallskip}
            \hline
         \end{tabular}
   \end{table}

\begin{figure*}
  \begin{center}
  $\begin{array}{c@{\hspace{0.1cm}}c@{\hspace{0.1cm}}c}

  \multicolumn{1}{l}{\mbox{\bf }} &
        \multicolumn{1}{l}{\mbox{\bf }} & \multicolumn{1}{l}{\mbox{\bf }}\\ [0cm]
  \epsfxsize=5.9cm
  \epsffile{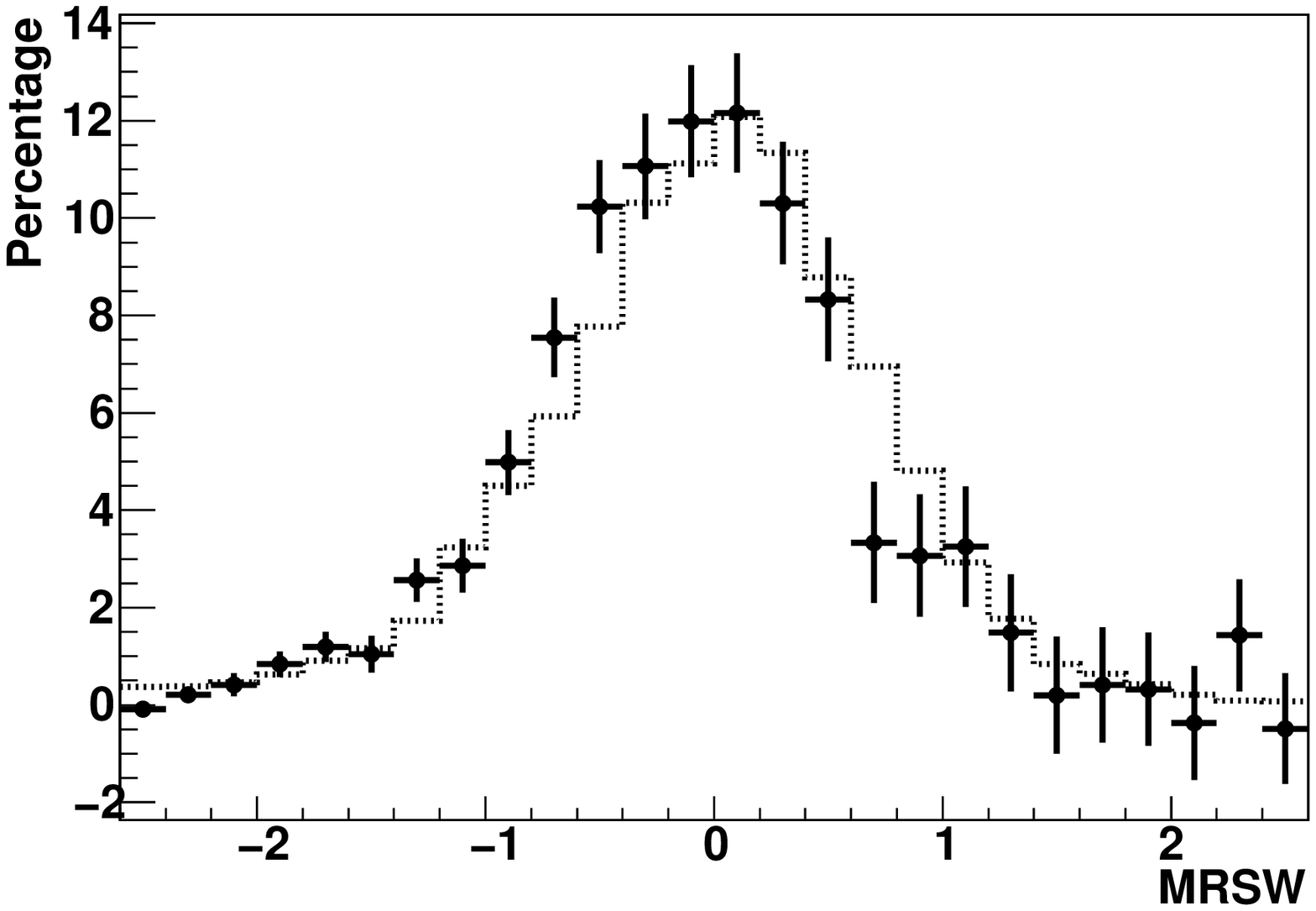} &
  \epsfxsize=5.9cm
  \epsffile{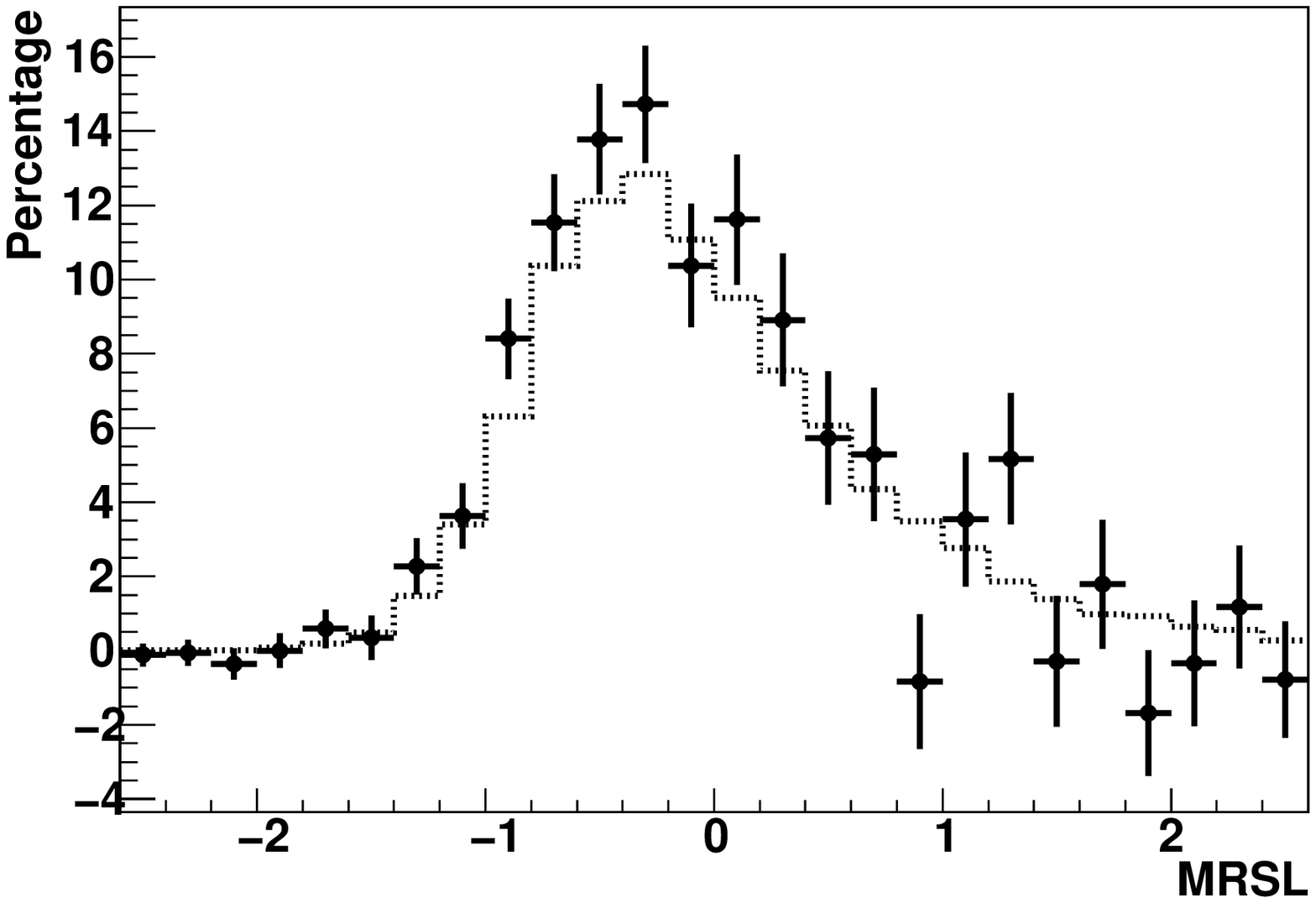} &
  \epsfxsize=5.9cm
  \epsffile{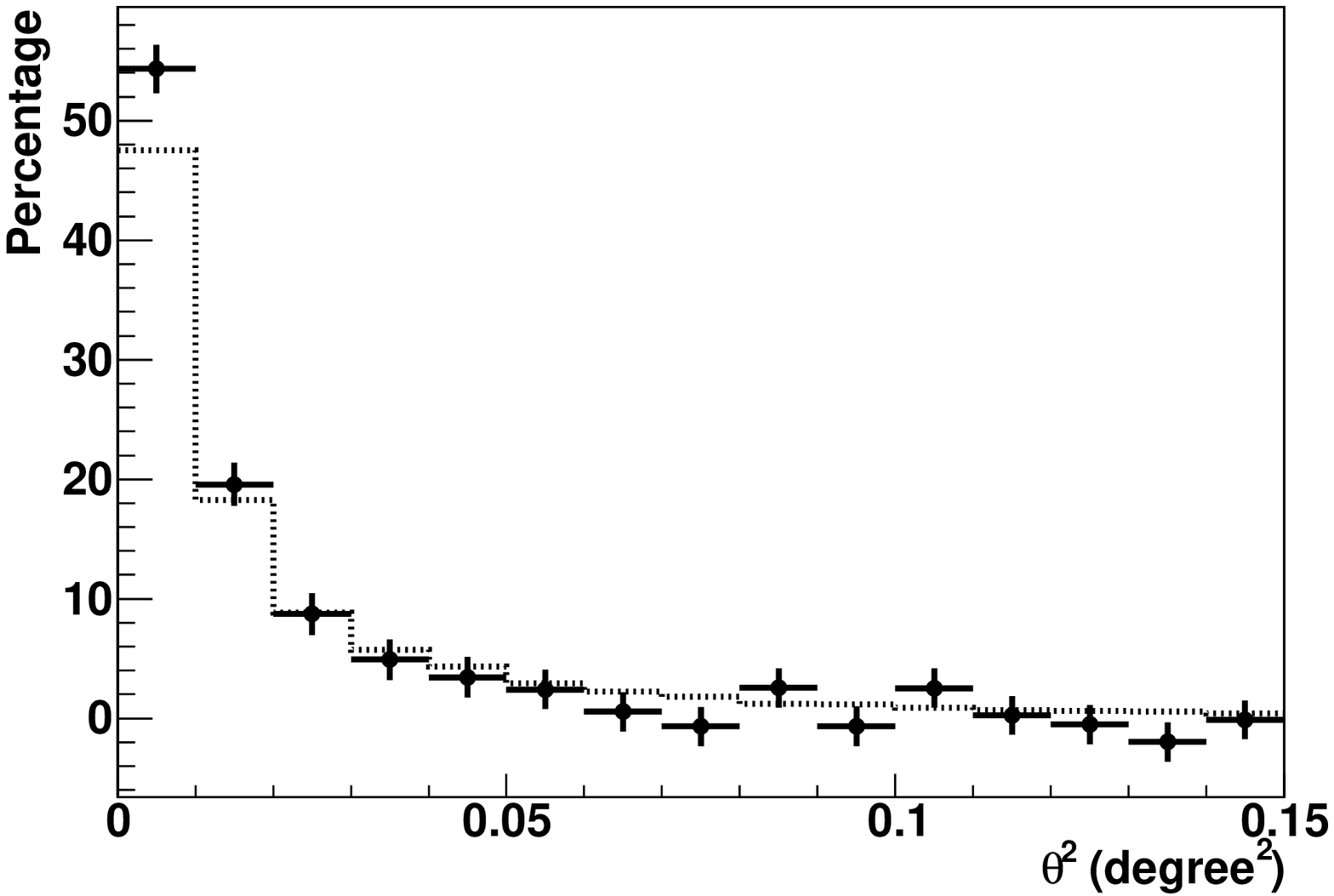} \\ [0.cm]
  \mbox{\bf (a)} & \mbox{\bf (b)} & \mbox{\bf (c)}
  \end{array}$	
  \end{center}
  \caption{The distributions of {\bf a)} mean reduced scaled width (MRSW), 
{\bf b)} mean reduced scaled length (MRSL), and {\bf c)} $\theta^{2}$ for the background subtracted signal after selection cuts 
(excluding both the MRSW and MRSL cuts in (a) and (b), and the $\theta^{2}$ cut in (c)) 
from observations of PKS 2155$-$304 (points) and similarly for Monte Carlo gamma-ray simulations
(spectral index = $3.35$, zenith angle = 20$^{\circ}$) in the two-telescope hardware stereo configuration (line).}	
  \label{adv_pars}
  \end{figure*}

Figure~\ref{adv_pars} shows the distributions of MRSW, MRSL, and $\theta^{2}$ for the observed excess 
after all cuts are applied (excluding both the MRSW and MRSL cuts in (a) and (b), and the $\theta^{2}$ cut in (c)), 
as well as the corresponding distributions for Monte Carlo gamma-ray simulations, for the entire two-telescope
hardware stereo data set.  The agreement is good and shows that the simulations accurately characterize the detector response 
and that the detector performs as expected.  In particular, the agreement of the $\theta^{2}$ distribution
demonstrates that the point spread function (width$<$0.1$^{\circ}$) is well understood. 

\section{Energy Spectrum}
\subsection{Technique}

\begin{table*}
\caption{The results of power law fits to the differential energy spectrum of PKS 2155$-$304 for each dark period. 
Shown are $I_o$, $\Gamma$, the $\chi^{2}$ and degrees of freedom (d.o.f) for the fit,
the integral flux above 1 TeV (I($>$1 TeV)), I($>$300 GeV), I($>E_{\mathrm{th}}$) where E$_{\mathrm{th}}$ is
the post-selection cuts energy threshold of the observations, and the percentage of these fluxes
relative to the respective values measured by H.E.S.S. from the Crab Nebula.
Only the statistical errors are shown.}
\label{spectra_results}
\centering
\begin{tabular}{r c c c c c c c c c c c c}
\hline\hline
\noalign{\smallskip}
Dark & I$_{o}$ & $\Gamma$ & $\chi^{2}$ (d.o.f.) & I($>$1 TeV) & Crab & I($>$300 GeV) & Crab & E$_{\mathrm{th}}$ & I($>$E$_{\mathrm{th}}$) & Crab \\
Period & $\left[ \frac{10^{-8}}{\mathrm{m}^2 \hspace{0.1cm} \mathrm{s} \hspace{0.1cm} \mathrm{TeV}} \right]$ & & & $\left[ \frac{10^{-9}}{\mathrm{m}^2 \hspace{0.1cm} \mathrm{s}} \right]$ & \% & $\left[ \frac{10^{-7}}{\mathrm{m}^2 \hspace{0.1cm} \mathrm{s}} \right]$ & \% & [GeV] & $\left[ \frac{10^{-7}}{\mathrm{m}^2 \hspace{0.1cm} \mathrm{s}} \right]$ & \% \\
\noalign{\smallskip}
\hline
\noalign{\smallskip}
07/2002 & 15.6$\pm$2.1 & 2.84$\pm$0.24 & 7.0 (7) & 85.1$\pm$16.1 & 43 & 7.77$\pm$1.63 & 56 & 300 & 7.77$\pm$1.63 & 56\\
10/2002 & 6.36$\pm$1.75 & 3.10$\pm$0.46 & 0.5 (4) & 30.4$\pm$10.6 & 15 & 3.79$\pm$1.63 & 28 & 300 & 3.79$\pm$1.63 & 28\\
06/2003 & 2.42$\pm$0.28 & 3.56$\pm$0.17 & 4.0 (5) & 9.45$\pm$1.27 & 4.8 & 2.06$\pm$0.37 & 15 & 260 & 2.97$\pm$0.59 & 17\\
07/2003 & 1.75$\pm$0.18 & 3.26$\pm$0.11 & 3.9 (7) & 7.73$\pm$0.87 & 3.9 & 1.18$\pm$0.15 & 8.6 & 170 & 4.27$\pm$0.74 & 12\\ 
08/2003 & 1.84$\pm$0.18 & 3.36$\pm$0.09 & 7.4 (7) & 7.78$\pm$0.82 & 3.9 & 1.34$\pm$0.16 & 9.7 & 170 & 5.13$\pm$0.80 & 15\\
09/2003 & 2.40$\pm$0.41 & 3.42$\pm$0.15 & 2.9 (6) & 9.91$\pm$1.79 & 5.0 & 1.82$\pm$0.38 & 13 & 160 & 8.35$\pm$2.27 & 22\\
\noalign{\smallskip}
\hline
\end{tabular}
\end{table*}

The energy of observed gamma rays, E$_{\mathrm{fit}}$, is calculated using the mean of the energies estimated for 
each telescope with a typical event resolution of $\sim$15$\%$.
Each of the individual energy estimates use the image size, impact parameter\footnote{The distance of the center of gravity
of the image from the center of the field of view is used in the mono-telescope configuration.} 
of the event, and the zenith angle of observation (Z) for each telescope.  The energy estimates are
based on results from Monte Carlo gamma-ray simulations.  
For the determination of the energy spectrum the data are logarithmically binned (8 bins per decade) in energy.
However, only bins for which the average bias in E$_{\mathrm{fit}}$ is less than 10\% are 
used, effectively placing a lower energy threshold on suitable events.  This is done to eliminate
any systematic effects that might arise because energy estimates
for events near the trigger threshold yield, on average, too large an energy.
The range of bins used varies for each individual run 
due to the zenith angle of observation and results in the bins at lower energies having differing amounts of exposure. 
For each event passing cuts the effective area, A$_{\mathrm{eff}}$(E$_{\mathrm{fit}}$,Z), is individually determined. 
The influence of the finite energy resolution has been taken into account in the effective areas which are fit 
as a function of Z and E$_{\mathrm{fit}}$ for each of the various observation configurations
similar to the method in ~\cite{Mohanty}.  For each bin, the sum of 1/ A$_{\mathrm{eff}}$, is
calculated for both on-source ($O_{i}$) and off-source ($B_{i}$) events
using only events from the appropriate runs. Each bin in the spectrum then contains:
\begin{equation}
\label{spec_point}
\begin{array}{c}
F_{i} = \frac{(O_{i} - B_{i})}{\Delta E_{i} t_{i}} \hspace{0.5ex} , 
\end{array}
\end{equation} 
\noindent
where $t_{i}$ and $\Delta E_{i}$ are the exposure time and width in energy respectively for that bin.  
As the collection area depends weakly on the spectral shape assumed, an iterative procedure is followed, 
where the initial spectral shape is assumed to be flat in $\nu F_{\nu}$ (differential power-law index=$2.0$).  
When deviations from this hypothesis are found the collection area is recalculated using the previous result 
and the spectrum is determined again.  This procedure continues 
until convergence (typically 1-2 iterations).  

\subsection{Dark Period Spectra}

\begin{figure*}
  \begin{center}
  $\begin{array}{c@{\hspace{0.5cm}}c}

  \multicolumn{1}{l}{\mbox{\bf }} &
        \multicolumn{1}{l}{\mbox{\bf }} \\ [0cm]
  \epsfxsize=8.7cm
  \epsffile{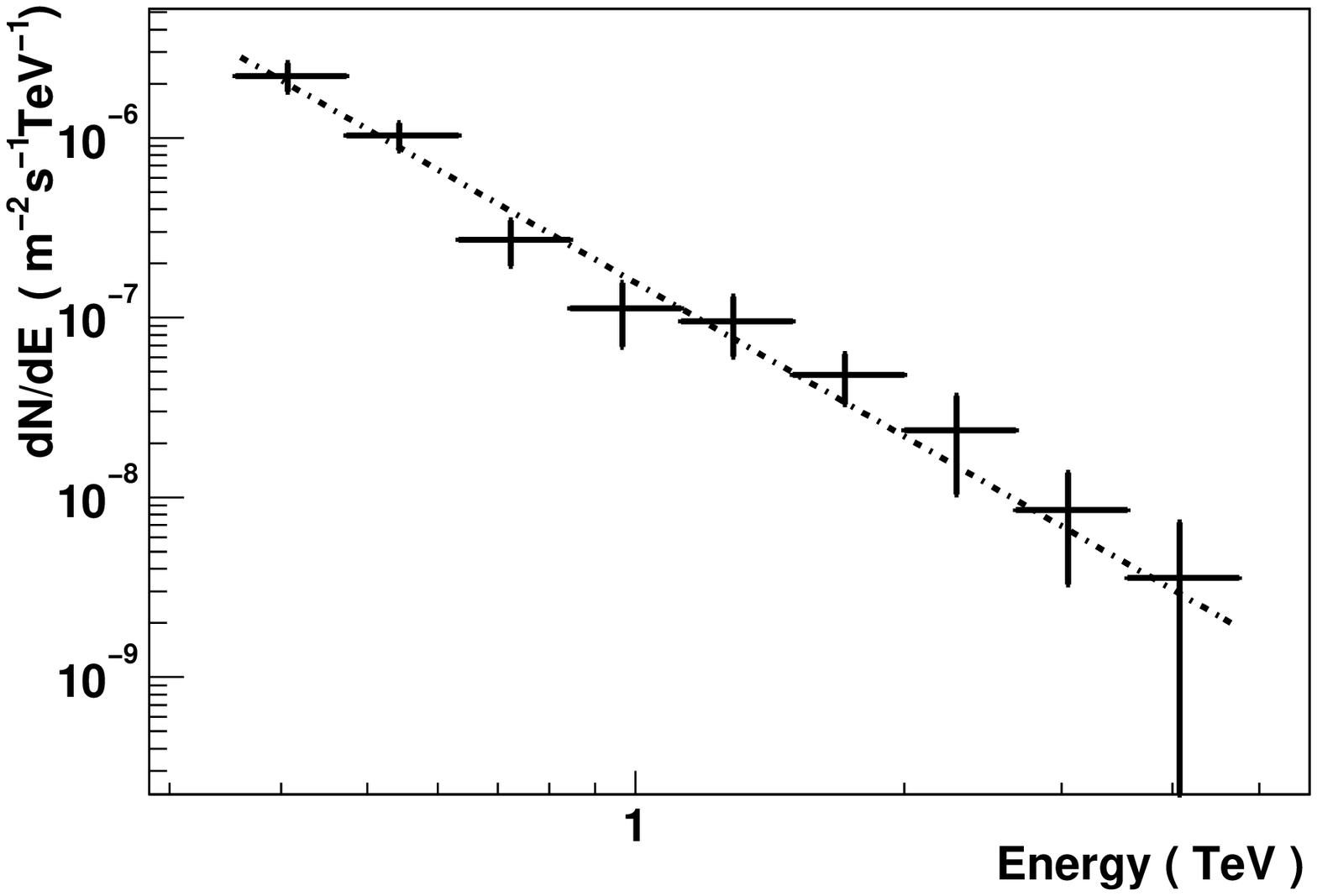} &
  \epsfxsize=8.7cm
  \epsffile{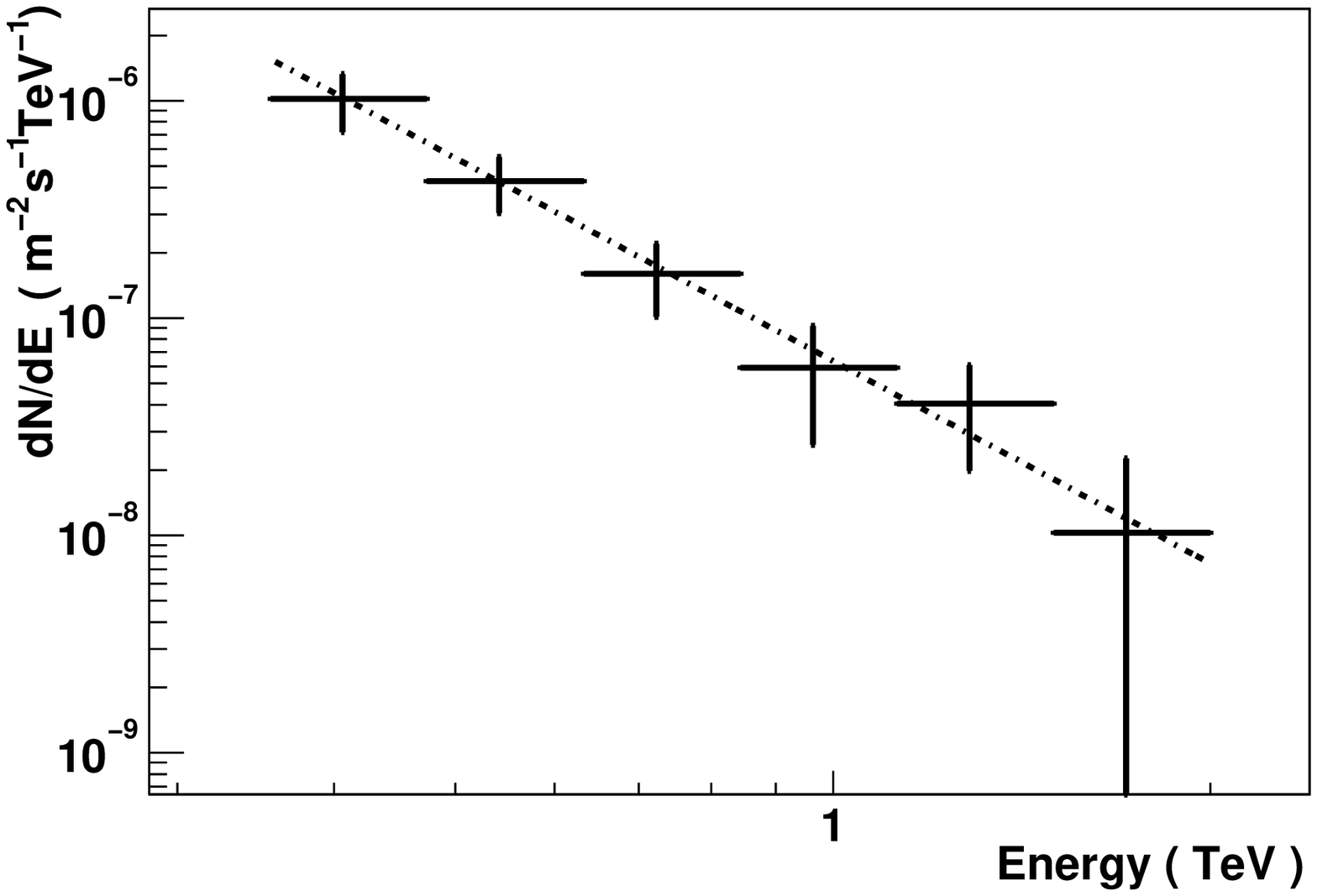} \\ [0.cm]
  \mbox{\bf (a)} & \mbox{\bf (b)} \\  [0.1cm]

  \epsfxsize=8.7cm
  \epsffile{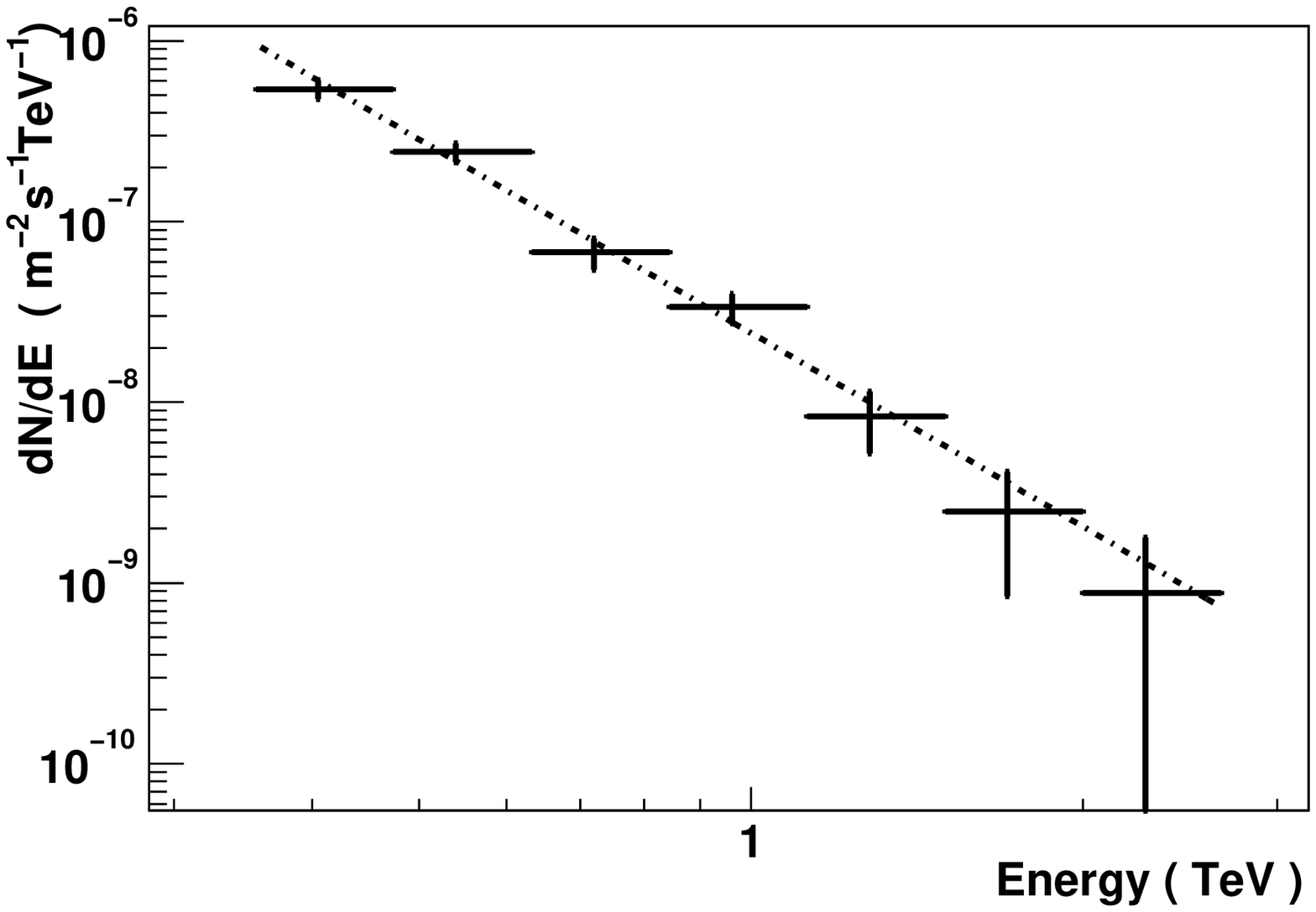} &
  \epsfxsize=8.7cm
  \epsffile{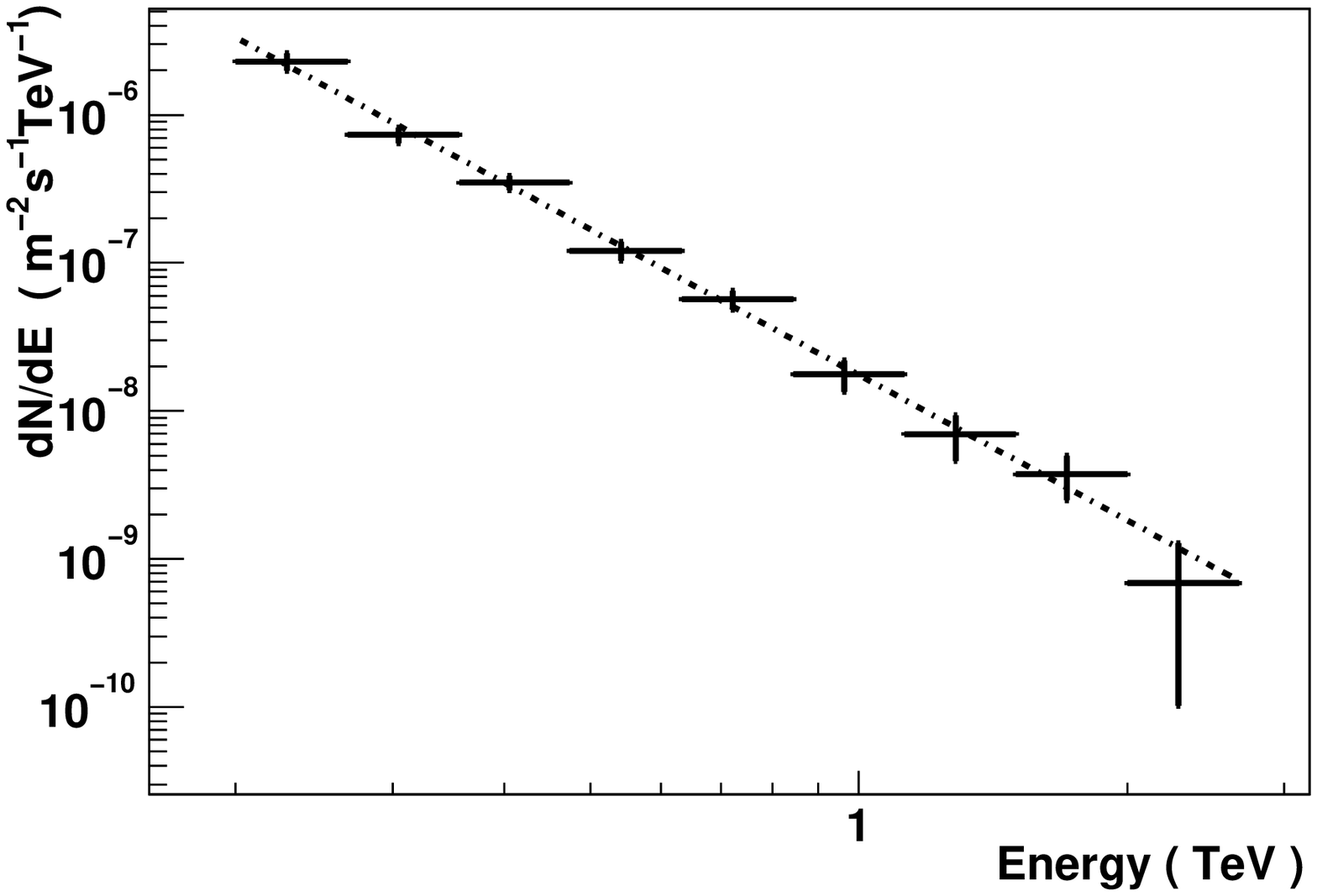} \\ [0.cm]
  \mbox{\bf (c)} & \mbox{\bf (d)}  \\ [0.2cm]

  \epsfxsize=8.7cm
  \epsffile{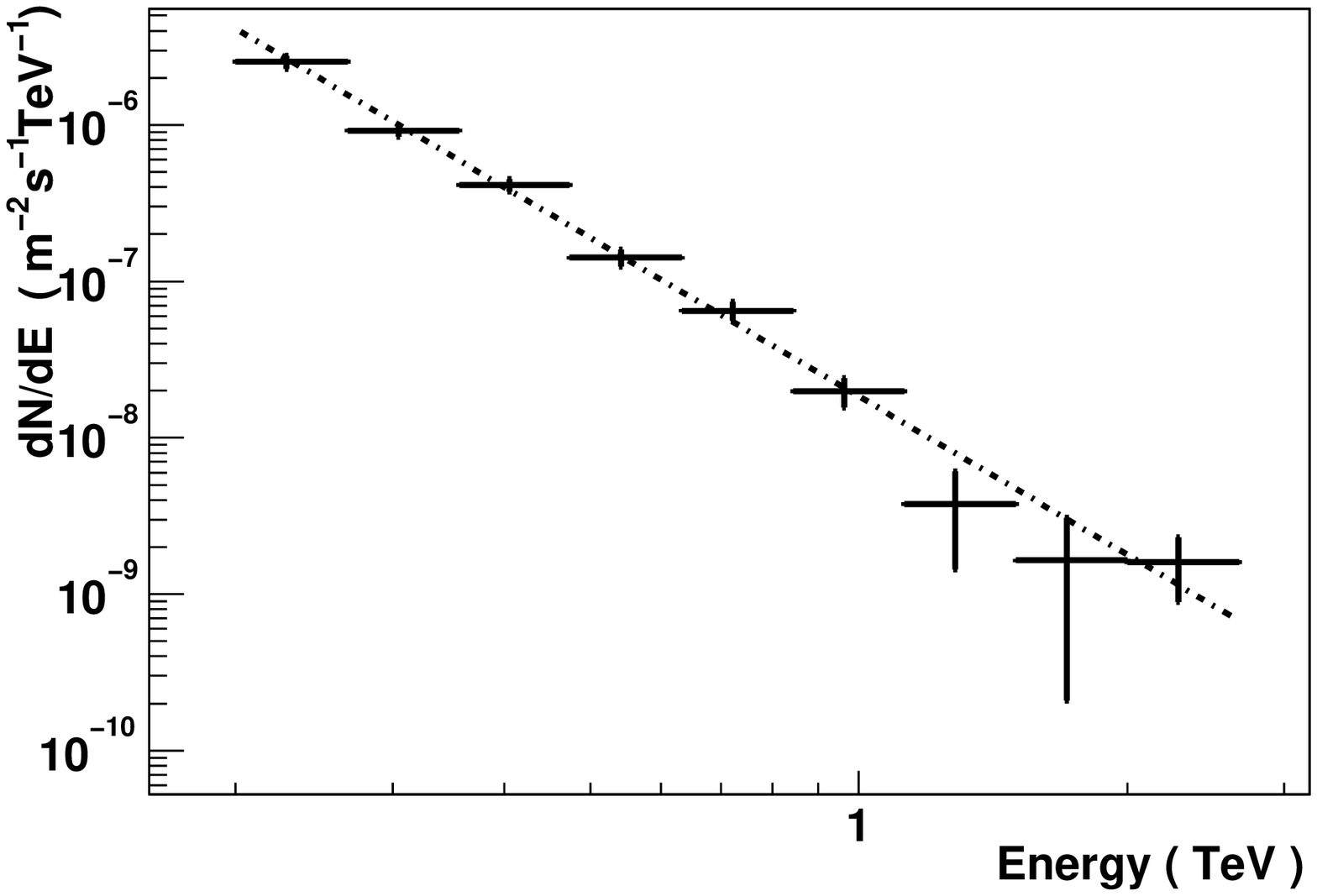} &
  \epsfxsize=8.7cm
  \epsffile{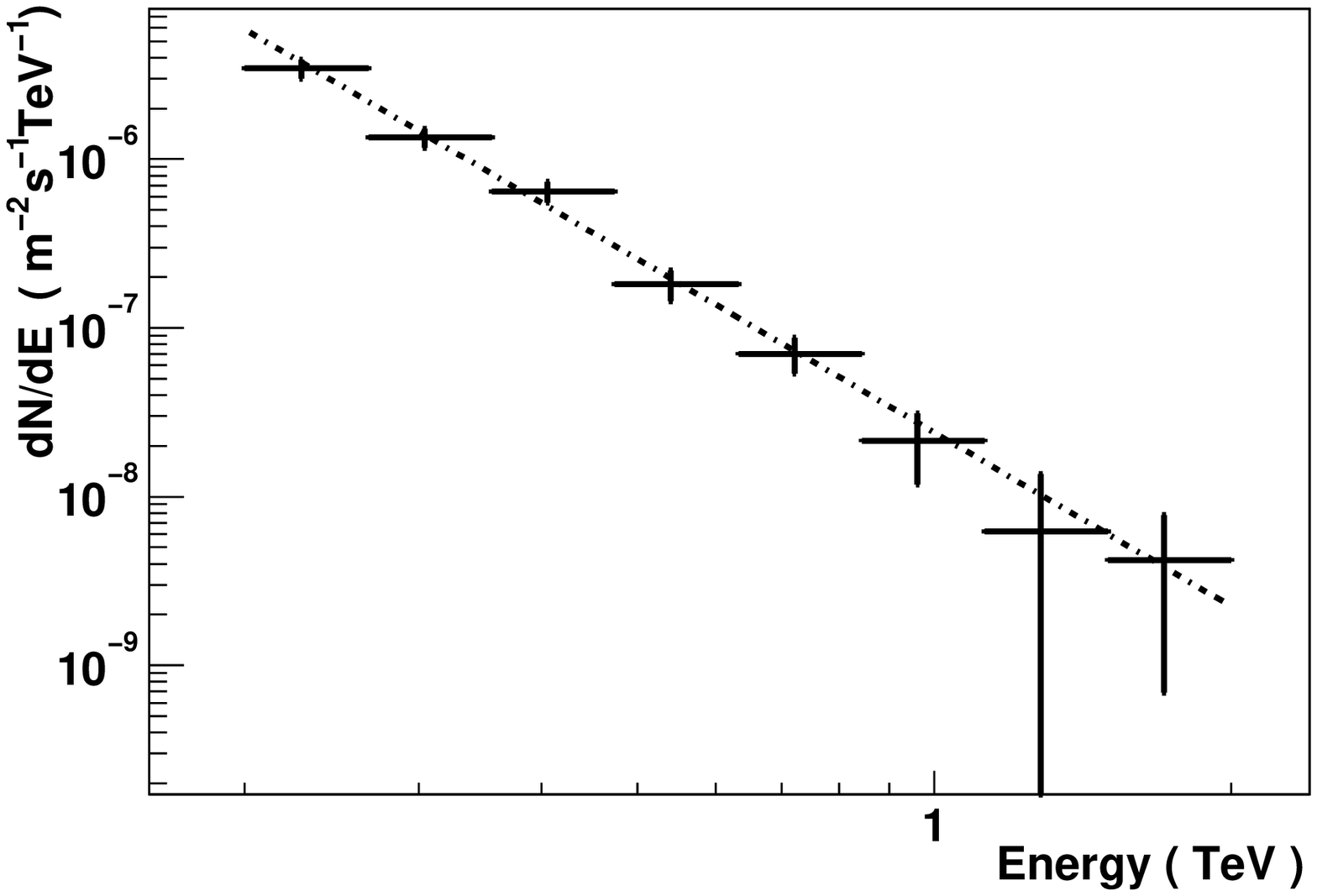} \\ [0.cm]
  \mbox{\bf (e)} & \mbox{\bf (f)}

  \end{array}$
                                                                                
  \end{center}
  \caption{The differential energy spectrum 
	measured from PKS 2155$-$304 for the dark periods of {\bf a)} July 2002, {\bf b)} October 2002, {\bf c)} June 2003,
	{\bf d)} July 2003, {\bf e)} August 2003, {\bf f)} September 2003.  
	The dashed lines represent the best $\chi^2$ fit of the data to a power law.}
  \label{monthly_spectra}
  \end{figure*}

Using the observed excess of gamma rays, a differential spectrum for each darkness period during which PKS 2155$-$304 was detected
is generated, and is shown in Figure~\ref{monthly_spectra}.  These spectra are fit to a power law:
\begin{equation}
\label{diff_flux}
\begin{array}{c}
\frac{dN}{dE} = I_{o} \hspace{0.5ex} \left(\frac{E}{E_{o}}\right)^{-\Gamma} \hspace{0.5ex} ,
\end{array}
\end{equation} 
\noindent
where $I_{o}$ is the differential flux normalization, $E_{\circ}$ is the energy at which the flux is normalized (1 TeV), and 
$\Gamma$ is the spectral index of the source.  The dashed lines in Figure~\ref{monthly_spectra} show the
best $\chi^2$ fit to the data of Equation (\ref{diff_flux}) for the corresponding spectrum.
Table~\ref{spectra_results} shows the best fit results, integral flux (above several thresholds) determined from the fit parameters,
and comparisons of these integral fluxes to those measured by H.E.S.S. from the Crab Nebula. 
Due to the steeper spectral index determined for PKS 2155$-$304 ($\Gamma_{\mathrm{crab}} = 2.63\pm0.05$), the percentage of 
the respective Crab fluxes in an individual darkness period is larger for lower thresholds.
Although the level of VHE emission from AGN is known to be variable, the integral flux above 300 GeV measured for
each of the observation periods shows reasonable agreement with
the result reported by \cite{durham1}, 
I($>$300 GeV) = (4.2$\pm$0.75$_{\mathrm{stat}}$$\pm$2.0$_{\mathrm{syst}}$)$\times$10$^{-7}$ m$^{-2}$ s$^{-1}$.
In addition to the method described above, an alternative forward-folding method, used by {\it CAT} ~\cite{CAT_spec}, 
was also applied to quantify the individual spectra and yields
very similar results to those shown in Table~\ref{spectra_results}. 

\subsection{Time-Averaged Spectrum}

  \begin{figure}
  \centering
  $\begin{array}{c}
  \epsfxsize=8.7cm
  \epsffile{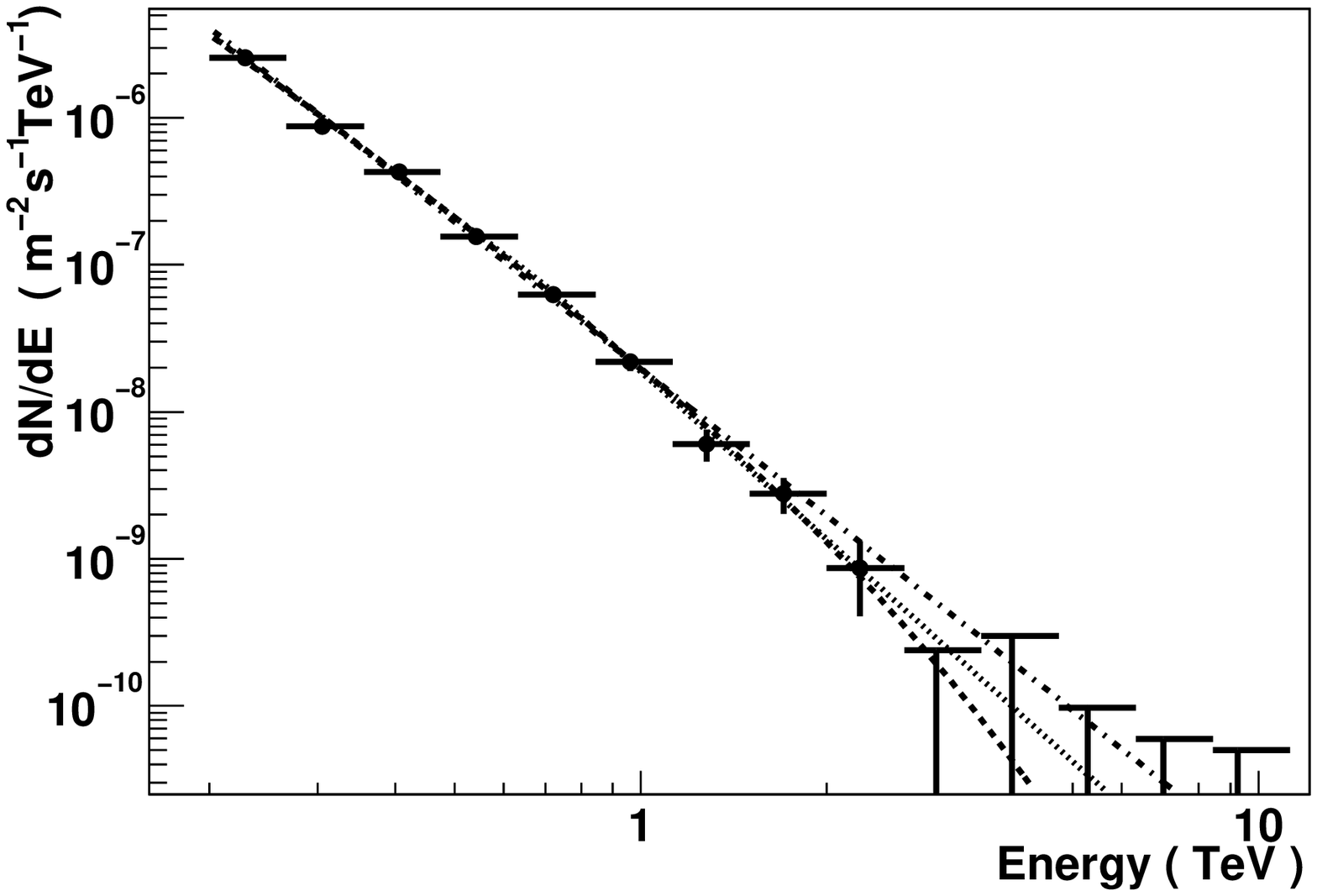} \\ [0cm]
  \epsfxsize=8.7cm
  \epsffile{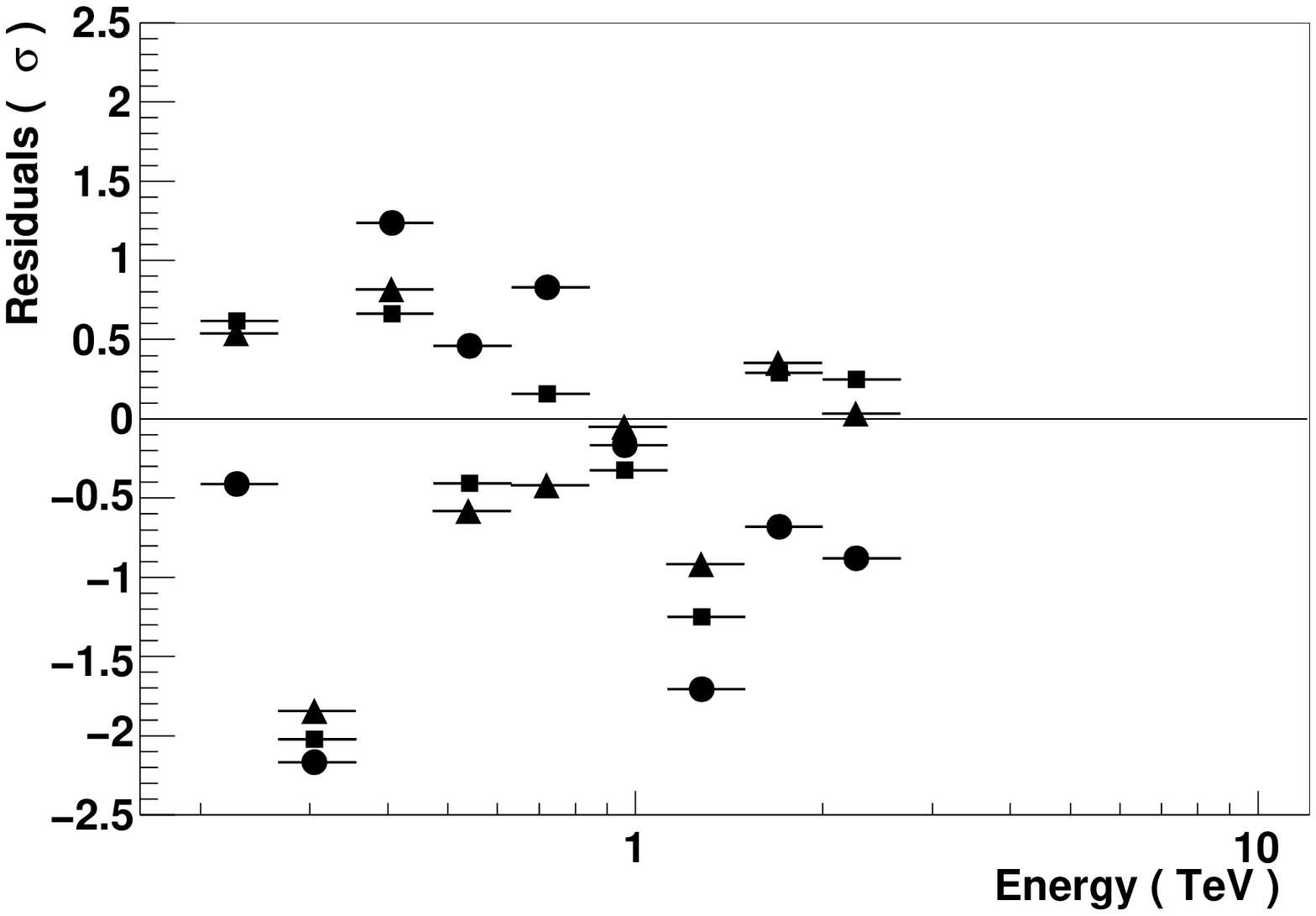} \\ [0cm]
  \end{array}$
  \caption{The time-averaged energy spectrum of PKS 2155$-$304 for the 2003 (stereo) data set (top).  The  90\%
confidence level upper limits are determined by the method of Helene (1983) and are not used in the fits.  The lines represent
the best $\chi^2$ fit to a power law (dot-dash), a power law with an exponential cutoff (dashed), and a broken power law (dotted). 
The bottom figure shows the residuals, (measurement $-$ fit) divided by the statistical error, of the best fits to a power law (circles), 
power law with exponential cutoff (squares), and broken power law (triangles).}
  \label{spectra_timeavg}
  \end{figure}

A time-averaged energy spectrum from all the stereo (2003) data is shown in Figure~\ref{spectra_timeavg}.
The mono-telescope data set is not included for two reasons: The average flux level in 2002 is considerably higher 
than the rest of the data set.  More importantly, the energy is more poorly determined for the mono-telescope data set and 
thus the determination of the spectrum is more susceptible to systematic errors.  A fit
of the stereo data set to a power law
yields $\Gamma$=$3.32\pm0.06$, $I_o$= (1.96$\pm$0.12) $\times$ 10$^{-8}$ m$^{-2}$ s$^{-1}$ TeV$^{-1}$, and
a $\chi^2$ of 10.8 for 7 degrees of freedom.  Only the statistical errors are presented.
Although the $\chi^2$ for the power law fits to the spectra for individual dark periods is reasonable,
the $\chi^2$ for the time-averaged spectrum corresponds to a probability of 0.148.  In addition, 
the power law fit to the time-averaged spectrum is above the 90\% upper limit on the differential flux at 3 TeV (not included in
the fit).  Since this information indicates that a power law is not a good characterization of the data, the energy
spectrum is fit to two alternative hypotheses.  The first is a power law with an exponential cutoff: 
   \begin{equation}
   \label{df_cut}
   \begin{array}{c}
	\frac{dN}{dE} = I_{\circ} \hspace{0.5 ex} \left(\frac{E}{E_{\circ}}\right)^{-\Gamma} \hspace{0.5 ex} 
	e^{-\frac{E}{E_{\mathrm{cut}}}} \hspace{0.5ex} ,
   \end{array}
   \end{equation}
\noindent
where $E_{\mathrm{cut}}$ is the cutoff energy, and $I_o$, $E_o$, and $\Gamma$ are as defined above.  
The fit yields an improved $\chi^2$ of 6.2 for 6 degrees of freedom (probability = 0.40), and the fit curve
falls below all upper limits at higher energies.  The best fit values are:
$\Gamma$=$2.90^{+0.21}_{-0.23}$, $I_o$= ($4.0^{+1.9}_{-1.2}$) $\times$ $10^{-8}$ m$^{-2}$ s$^{-1}$ TeV$^{-1}$, 
and $E_{\mathrm{cut}}$ = $1.4^{+0.8}_{-0.7}$ TeV.  The large errors (only statistical errors are shown) on the fit values are 
the result of strong correlations between the parameters used to characterize the spectral shape. Although the $\chi^2$ resulting from the
fit is improved, it cannot be concluded that the fit is significantly better since an F-test shows there is a probability of only 0.92 
that the power law with an exponential cutoff is a better characterization of the data.

The second hypothesis is a broken power law, which follows Equation (\ref{diff_flux}) for energies less than $E_{\mathrm{break}}$, and the following
for energies greater than $E_{\mathrm{break}}$:
   \begin{equation}
   \label{broken_pwr_law}
   \begin{array}{c}
	\frac{dN}{dE} = I_{\circ} \hspace{0.5 ex} \left(\frac{E_{\mathrm break}}{E_{\circ}}\right)^{(\Gamma_{2} - \Gamma_{1})} \left(\frac{E}{E_{\circ}}\right)^{-\Gamma_{2}} \hspace{0.5ex},
   \end{array}
   \end{equation}
\noindent
where $E_{\mathrm{break}}$ is the energy of the break point in TeV, $\Gamma_{1}$ and $\Gamma_{2}$
are the spectral indices,
and $I_o$ and $E_o$ are as defined above. 
The fit yields a $\chi^2$ of 5.1 for 5 degrees of freedom (probability = 0.40), 
and the fit curve does not fall below the upper limit at 3 TeV.  
The best fit values are:
$\Gamma_{1}$=$3.15^{+0.10}_{-0.12}$, 
$I_o$= ($2.4^{+0.4}_{-0.3}$) $\times$ $10^{-8}$ m$^{-2}$ s$^{-1}$ TeV$^{-1}$, 
$\Gamma_{2}$=$3.79^{+0.46}_{-0.27}$, 
and $E_{\mathrm{break}}$ = $0.7\pm0.2$ TeV.  
Only the statistical errors are shown. An F-test shows there is a probability of 0.85 that 
the broken power law is a better characterization of the data than the simple power law and again
it cannot be concluded that the fit is significantly better.

The possible presence of features, such as an exponential cutoff or break, in the energy spectrum could be the result of
absorption of TeV gamma rays by the extragalactic infrared background.  However, it may also be intrinsic to the 
source spectrum, or the result of absorption of photons close to the source, or both.  The distinction can only 
be made using a larger sample of TeV-emitting AGN at different redshifts, or perhaps by detailed modelling of the broad-band spectrum
using data gathered simultaneously at other wavelengths.

\section{Flux Variability}

\subsection{Monthly Time Scale}

  \begin{figure}
  \begin{center}
  $\begin{array}{c}
  \epsfxsize=8.7cm
  \epsffile{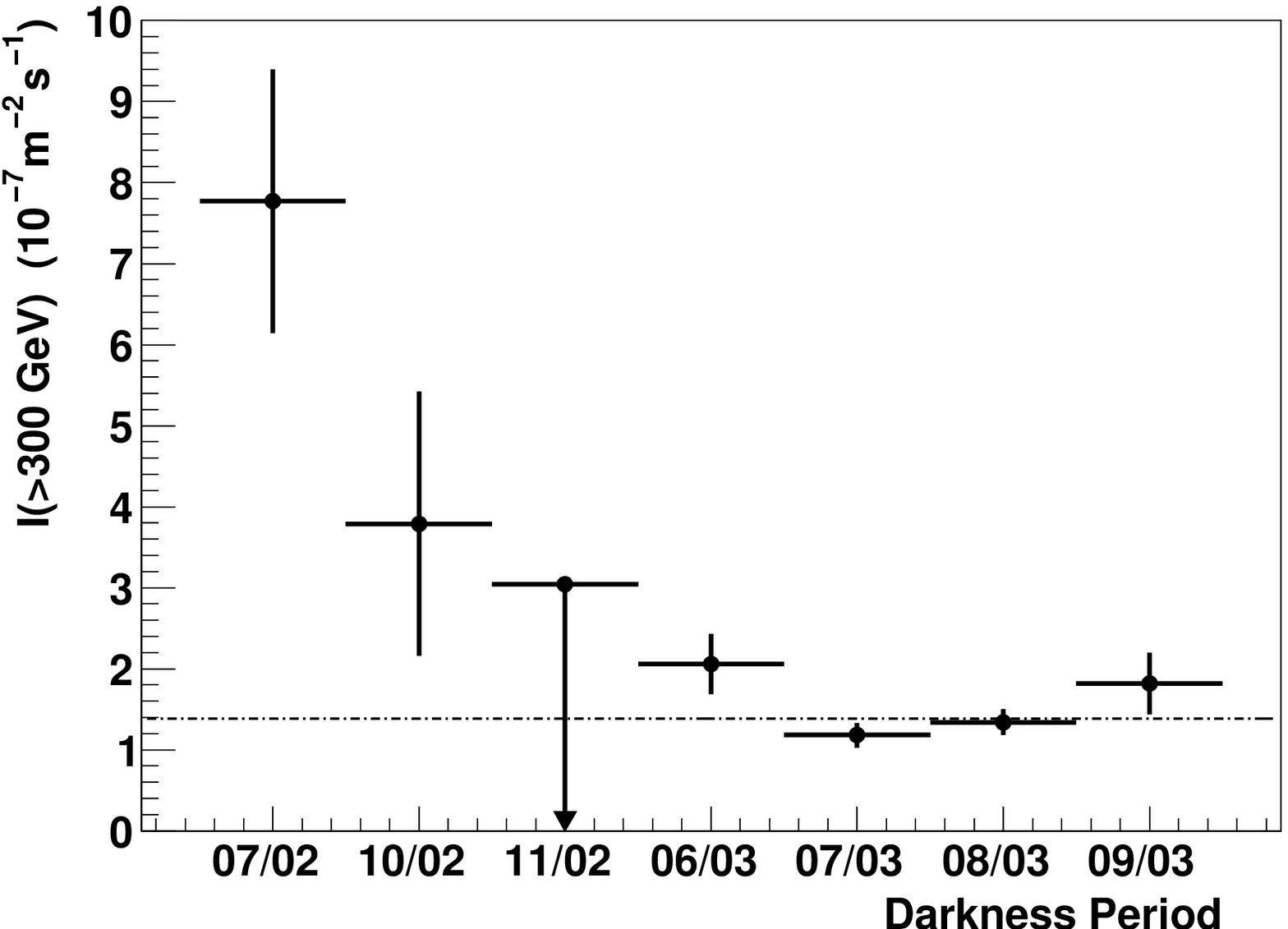} \\ [0cm]
  \epsfxsize=8.7cm
  \epsffile{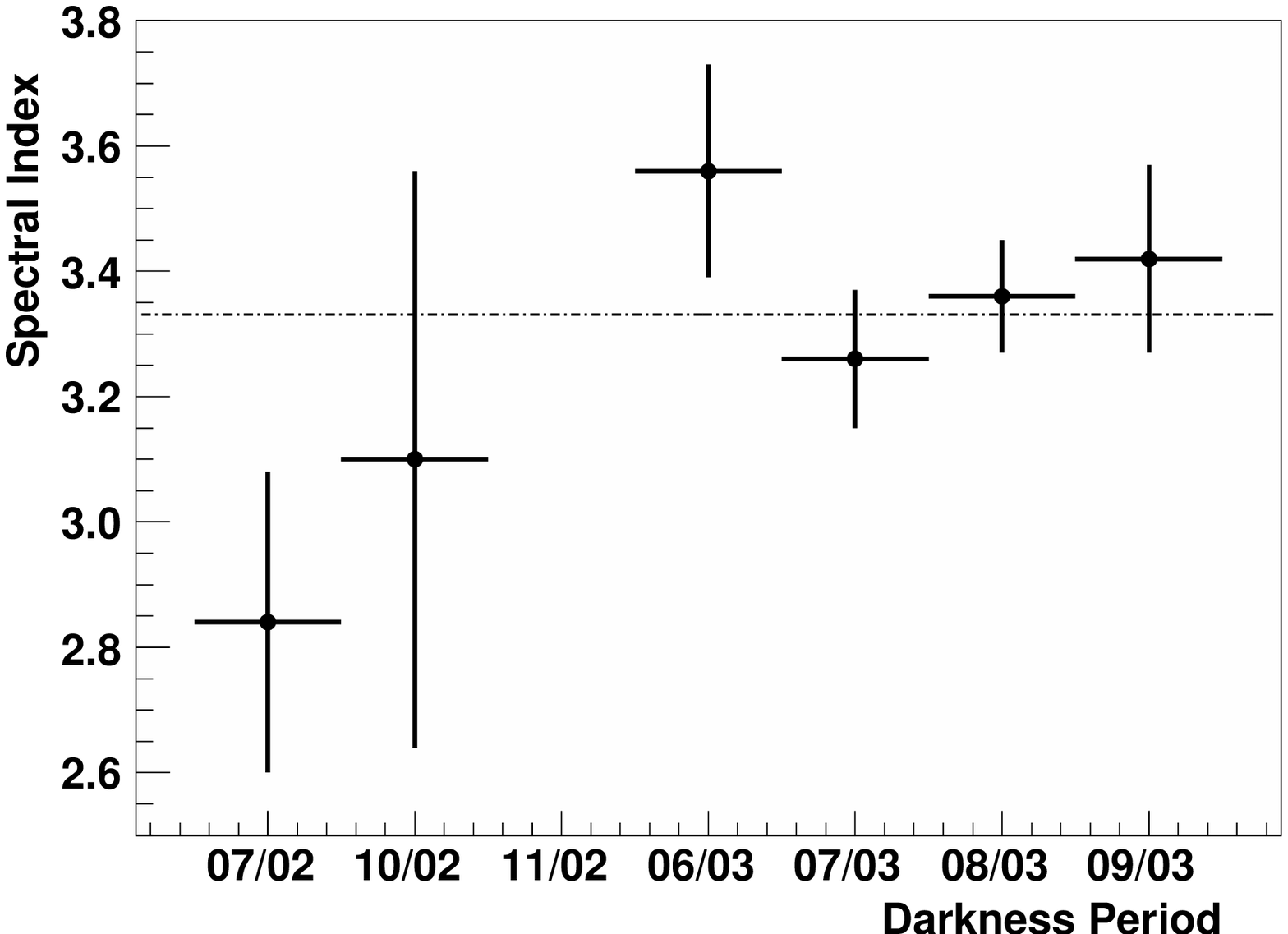} \\ [0cm]
  \end{array}$
  \end{center}
  \caption{The integral flux above 300 GeV and spectral index versus dark period determined from H.E.S.S. observations
of PKS 2155$-$304.  The dashed line represents the best $\chi^2$ fit of the data points to a constant.  Only the statistical errors are shown.}
  \label{fluxindvstime}
  \end{figure}
                   
Figure~\ref{fluxindvstime} shows the integral flux greater than 300 GeV (top) and fit spectral index (bottom) of PKS 2155$-$304 for each
dark period.  The dashed line in each plot represents the results from a $\chi^2$ fit of the data to a constant.
The fit of I($>$300 GeV) versus observation period yields a $\chi^2$ of 24.0
for 5 degrees of freedom, corresponding to a $2\times10^{-4}$ $\chi^2$ probability. The data are 
clearly inconsistent with a constant flux.  Although only the
statistical errors are used in the fit, the conclusion is unchanged even if systematic errors ($\sim$20$\%$ for the integral flux) are included. 
Since PKS 2155$-$304 was not detected in November 2002, a 99\% upper limit of I($>$300 GeV) = 3.0$\times10^{-7}$ m$^{-2}$ s$^{-1}$ 
was determined by the method of ~\cite{helene} assuming the 2003 time-averaged spectral index, $\Gamma=3.32$. The upper limit, shown in
Figure~\ref{fluxindvstime}, is below the flux values from earlier in 2002, but higher than the average flux for the entire data set.

The fit of the spectral index versus period to a constant yields a $\chi^2$ of 7.1 for 5 
degrees of freedom, corresponding to a 0.21 $\chi^2$ probability, 
which is marginally consistent with being constant.  Since the fit data points include only the 
statistical errors, the $\chi^2$ will be smaller with the inclusion of 
systematic errors ($\sim$0.1 for the spectral index).  Therefore, there is no evidence
that the spectral index of PKS2155$-$304 varies with time.  However, the evidence also does not rule out
possible temporal variability in the spectral index during the aforementioned large changes in flux.

A correlation between harder spectra and higher flux values might be expected for PKS 2155$-$304, since 
this trend was demonstrated for Mkn 421 ~\cite{HEGRA_421} and Mkn 501 (\cite{CAT_501}; ~\cite{HEGRA_501b}).
While Figure~\ref{fluxindvstime} is suggestive of such a
correlation, the inference is mainly drawn from the July 2002 data point, 
taken during the first month of H.E.S.S. operations where only one telescope was
operational, and thus the fit spectrum is most likely to have larger systematic errors.  
In addition, the spectral index was
shown to be marginally consistent with being constant.  Therefore, further observations
with the more sensitive complete 
H.E.S.S. array are needed to make a definitive statement regarding any correlation.

Given the large significance observed in each month of data taking, a variability search on a monthly time scale 
represents a rather coarse temporal binning of the data.  Therefore studies were performed 
on the data to determine if the VHE flux from PKS 2155$-$304 is also variable on shorter time scales. 

\subsection{Nightly Time Scale}

  \begin{figure*}
  \begin{center}
  $\begin{array}{c@{\hspace{0.5cm}}c}

  \multicolumn{1}{l}{\mbox{\bf }} &
        \multicolumn{1}{l}{\mbox{\bf }} \\ [0cm]
  \epsfxsize=8.7cm
  \epsffile{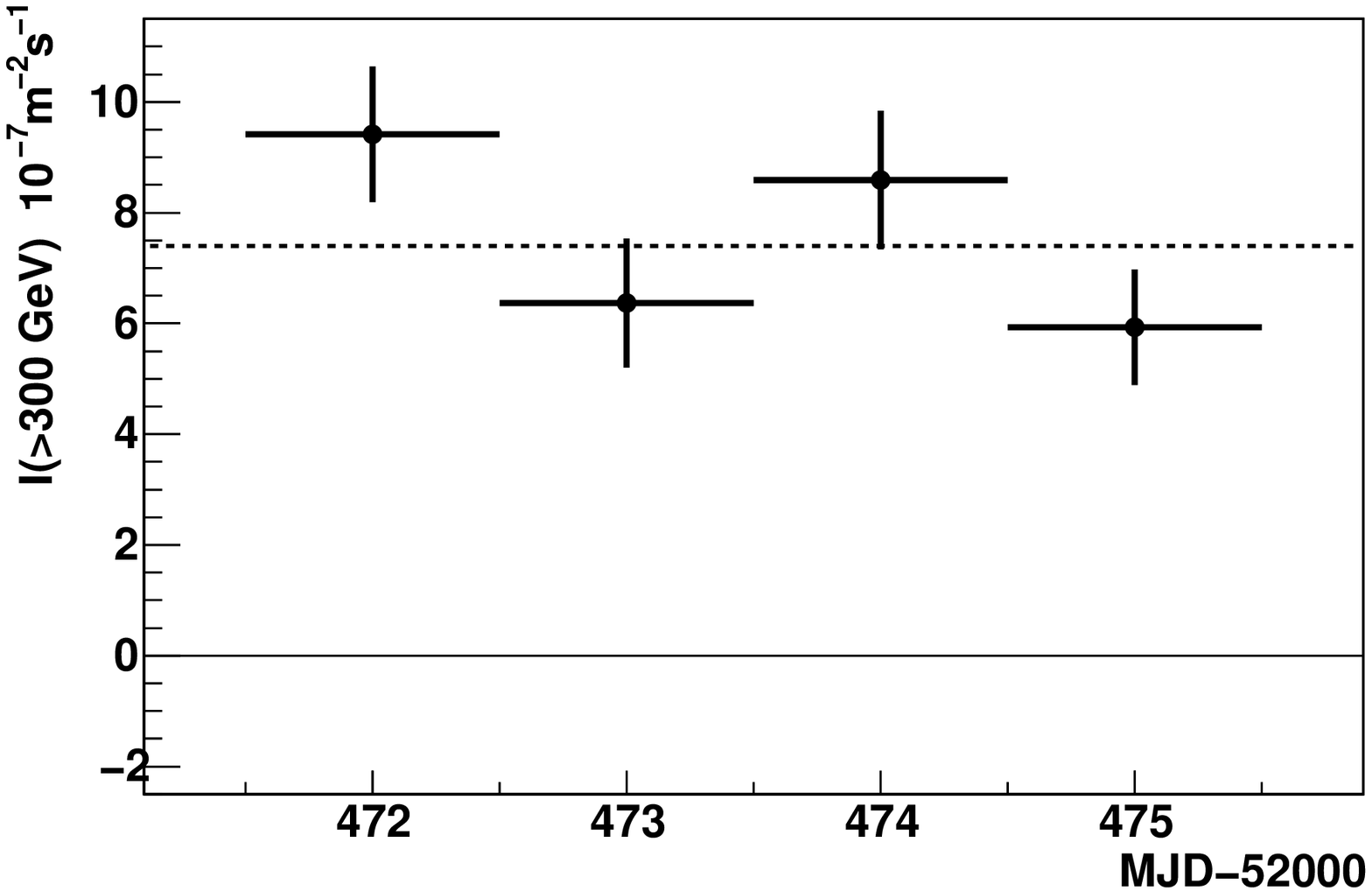} &
  \epsfxsize=8.7cm
  \epsffile{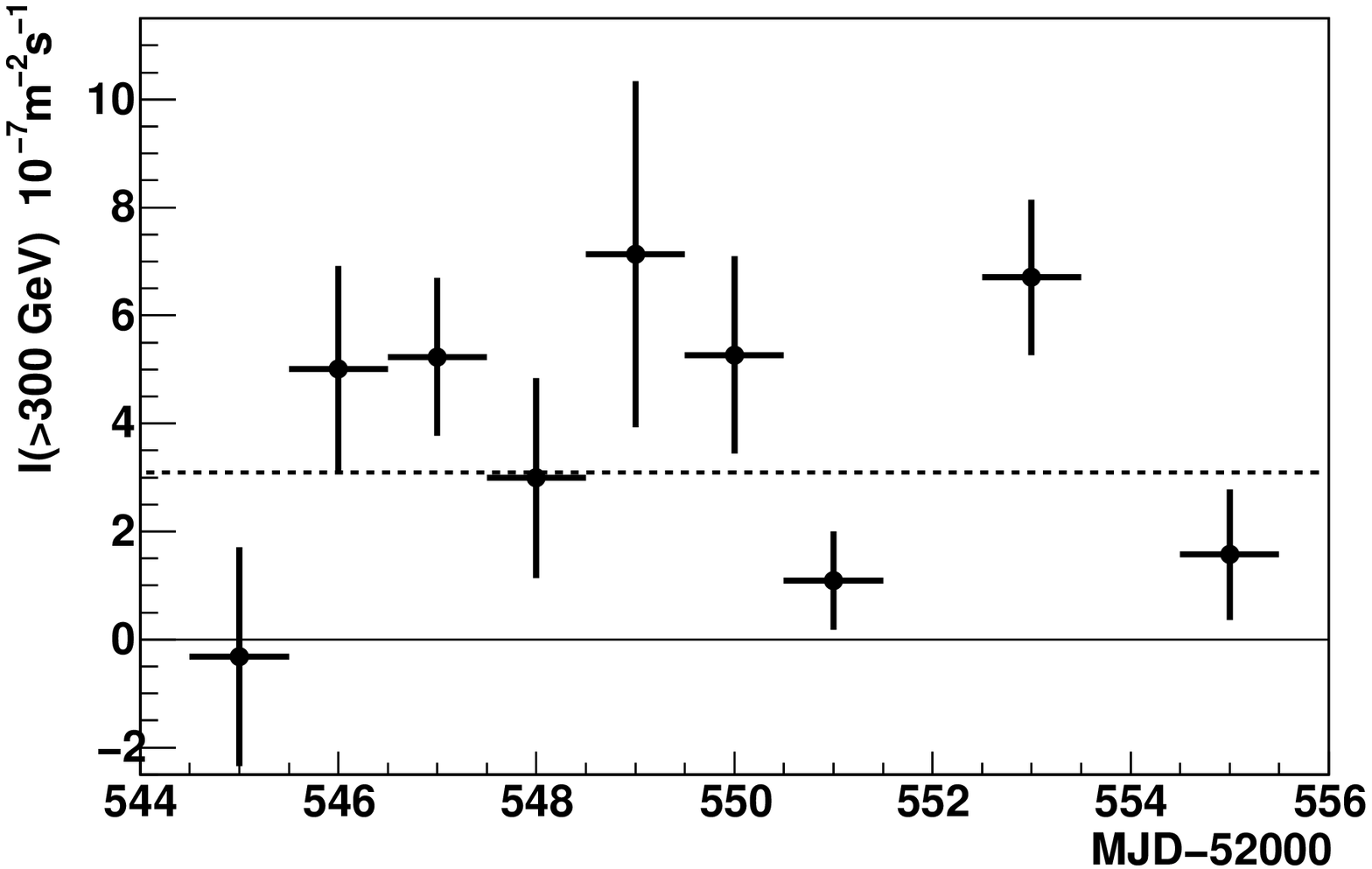} \\ [0.cm]
  \mbox{\bf (a)} & \mbox{\bf (b)} \\  [0.cm]

  \epsfxsize=8.7cm
  \epsffile{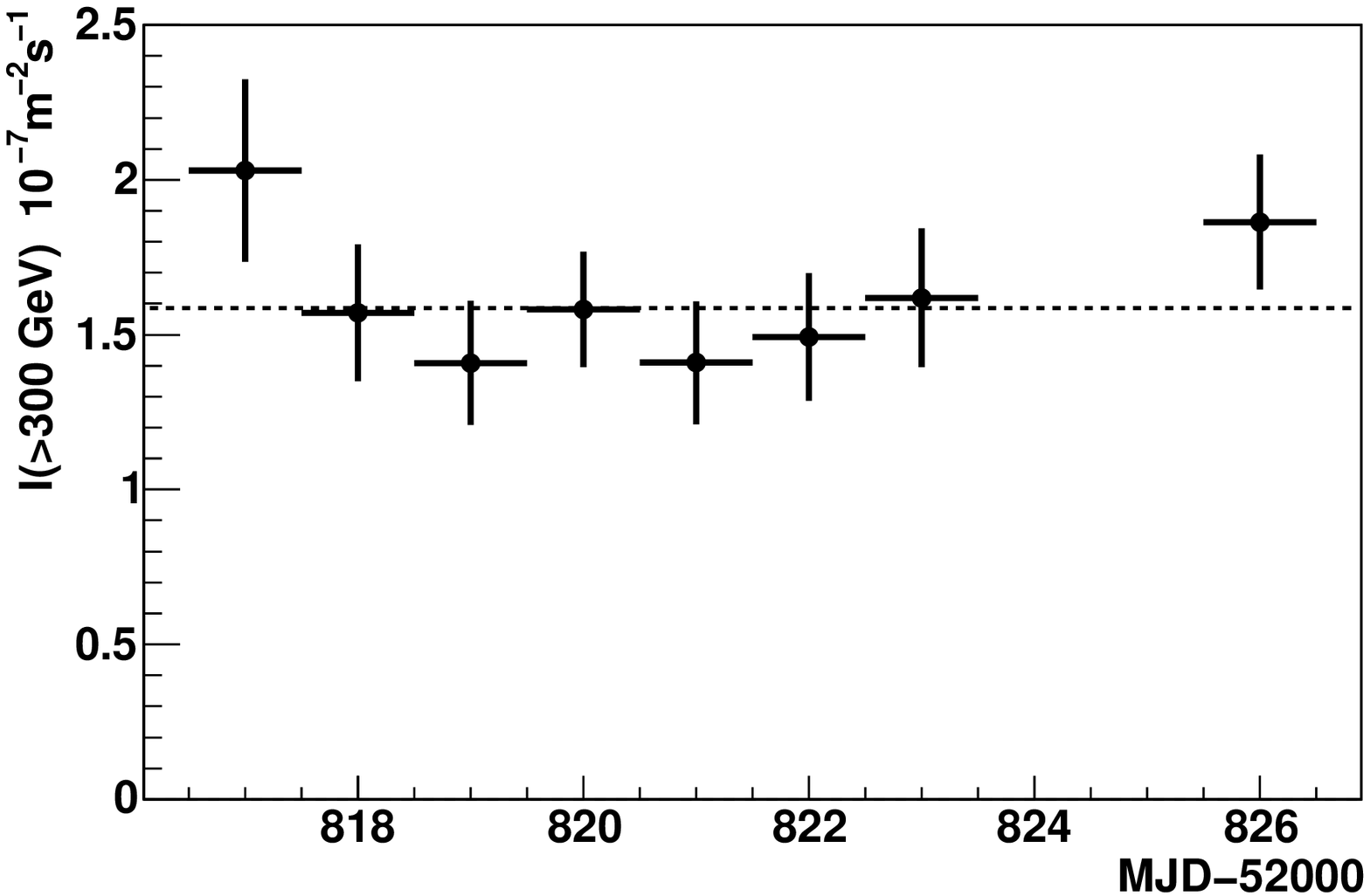} &
  \epsfxsize=8.7cm
  \epsffile{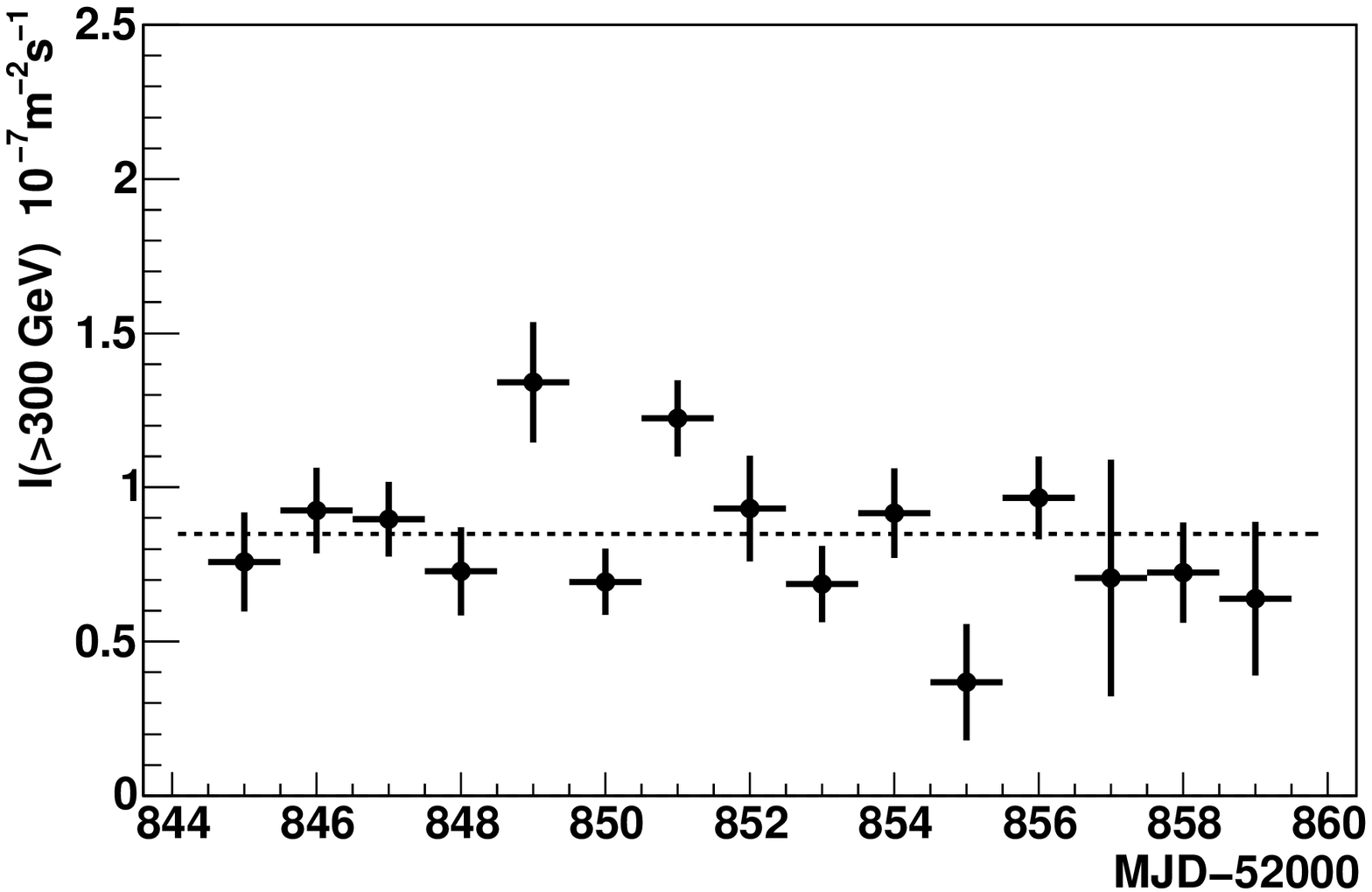} \\ [0.cm]
  \mbox{\bf (c)} & \mbox{\bf (d)}  \\ [0.cm]

  \epsfxsize=8.7cm
  \epsffile{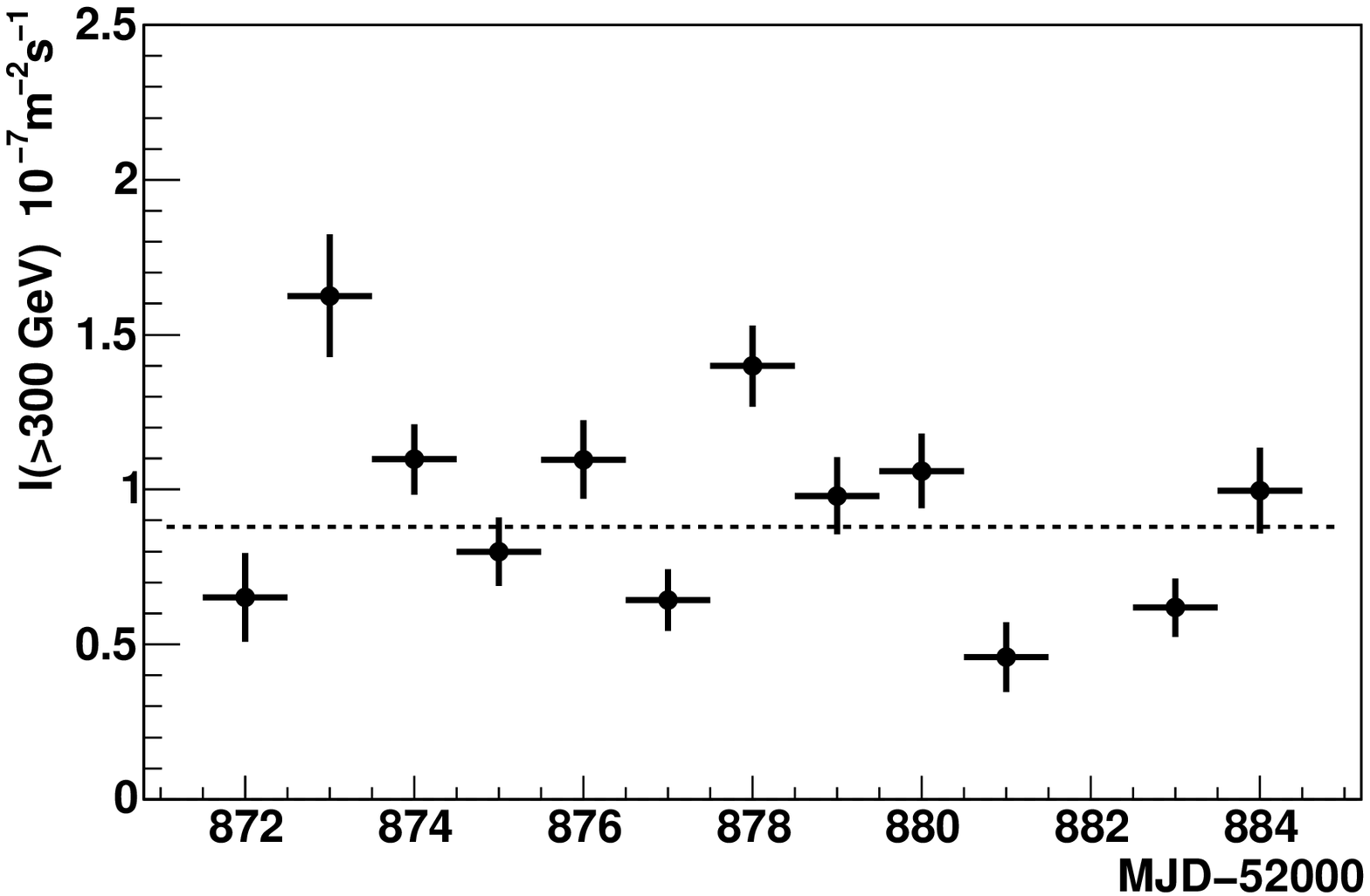} &
  \epsfxsize=8.7cm
  \epsffile{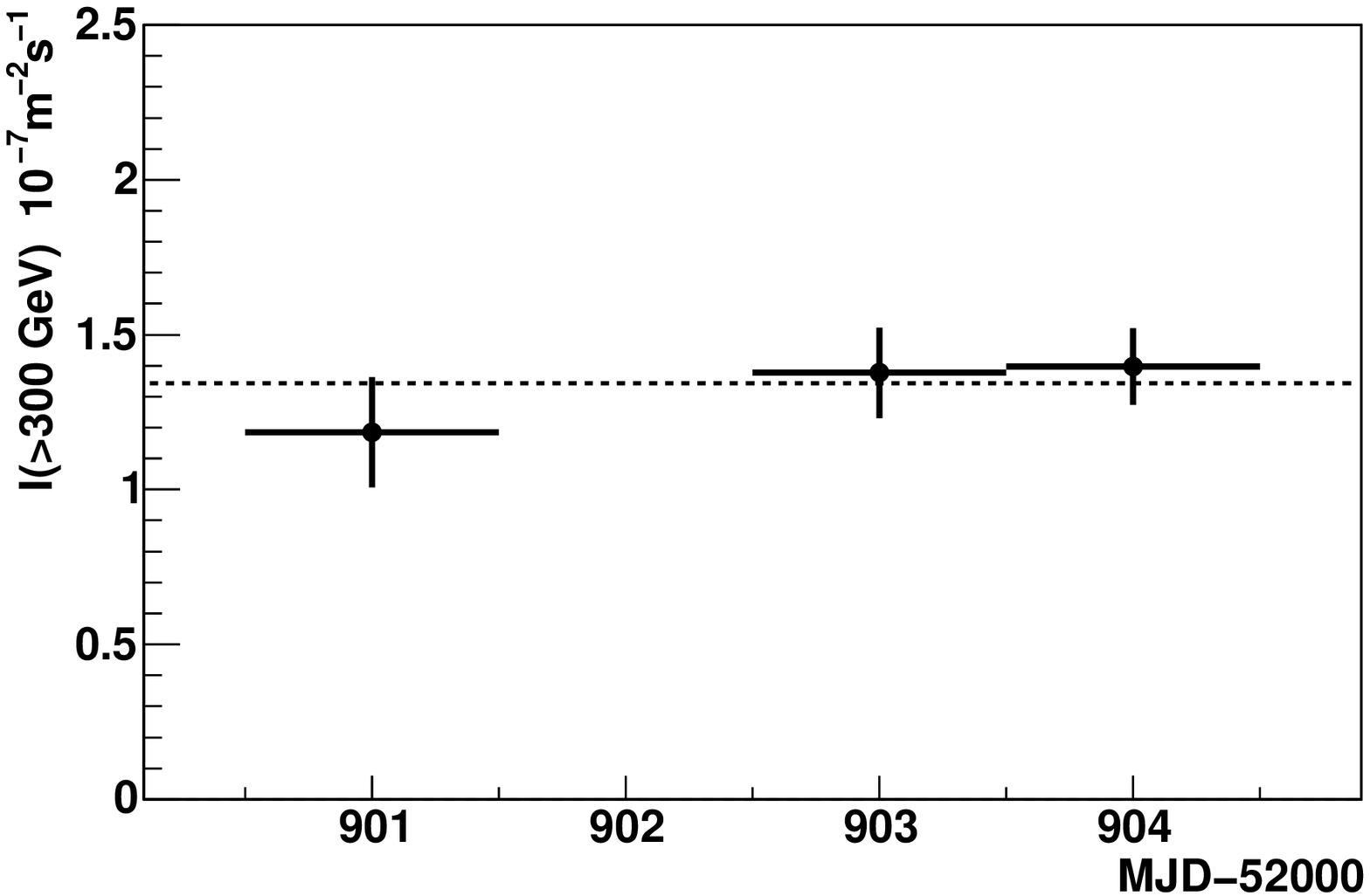} \\ [0.cm]
  \mbox{\bf (e)} & \mbox{\bf (f)}

  \end{array}$
                                                                                
  \end{center}
  \caption{The nightly integral flux greater than 300 GeV versus time 
	measured from PKS 2155$-$304 in {\bf a)} July 2002, {\bf b)} October 2002, {\bf c)} June 2003,
	{\bf d)} July 2003, {\bf e)} August 2003, {\bf f)} September 2003.  
	Only the statistical errors are shown.
	The dashed lines represent the best $\chi^2$ fit of the data to a 
	constant for each darkness period.}
  \label{nightlyfluxvstime}
  \end{figure*}
   
The integral flux greater than 300 GeV measured from PKS 2155$-$304 is calculated for each night of observations and is shown in
Figure~\ref{nightlyfluxvstime}.  The exposure is different in the various nights and
the spectral index assumed in the flux calculation is the value determined from the power law fit
to the time-averaged spectra for 2003 ($\Gamma=3.32$).  This value is used since there is
no strong evidence for variability in the spectral index.  Table~\ref{variability_results} lists the results of the fit of
the nightly fluxes during each month to a constant.  The poor $\chi^2$ values
for some of the dark periods show that variability in the flux of VHE gamma-rays from PKS 2155$-$304 clearly occurs on time scales of days. 
However, the June and September 2003 results also show that there are periods of steady flux as well.

\begin{table}[h]
\caption{The $\chi^2$, degrees of freedom (d.o.f.), and $\chi^2$ probability, from
the best fits to a constant of the nightly integral flux greater than 300 GeV observed from PKS 2155$-$304 versus 
time for each dark period. }
\label{variability_results}
\centering 
\begin{tabular}{r c c c}
\hline\hline
\noalign{\smallskip}
Dark Period & $\chi^{2}$ & d.o.f. & P($>\chi^{2}$)\\ 
\noalign{\smallskip}
\hline
\noalign{\smallskip}
07/2002 & 6.4 & 3 & 9.5$\times10^{-2}$ \\
10/2002 & 21.8 & 8 &  5.4$\times10^{-3}$\\
06/2003 & 5.7 & 7 & 0.58 \\
07/2003 & 30.1 & 14 & 7.4$\times10^{-3}$\\
08/2003 & 74.9 & 12 & 3.7$\times10^{-11}$\\
09/2003 & 1.0 & 2 & 0.60\\
\noalign{\smallskip}
\hline
\end{tabular}
\end{table}

\subsection{Sub-hour Time Scale}

  \begin{figure}
  \begin{center}
  $\begin{array}{c}

  \epsfxsize=8.7cm
  \epsffile{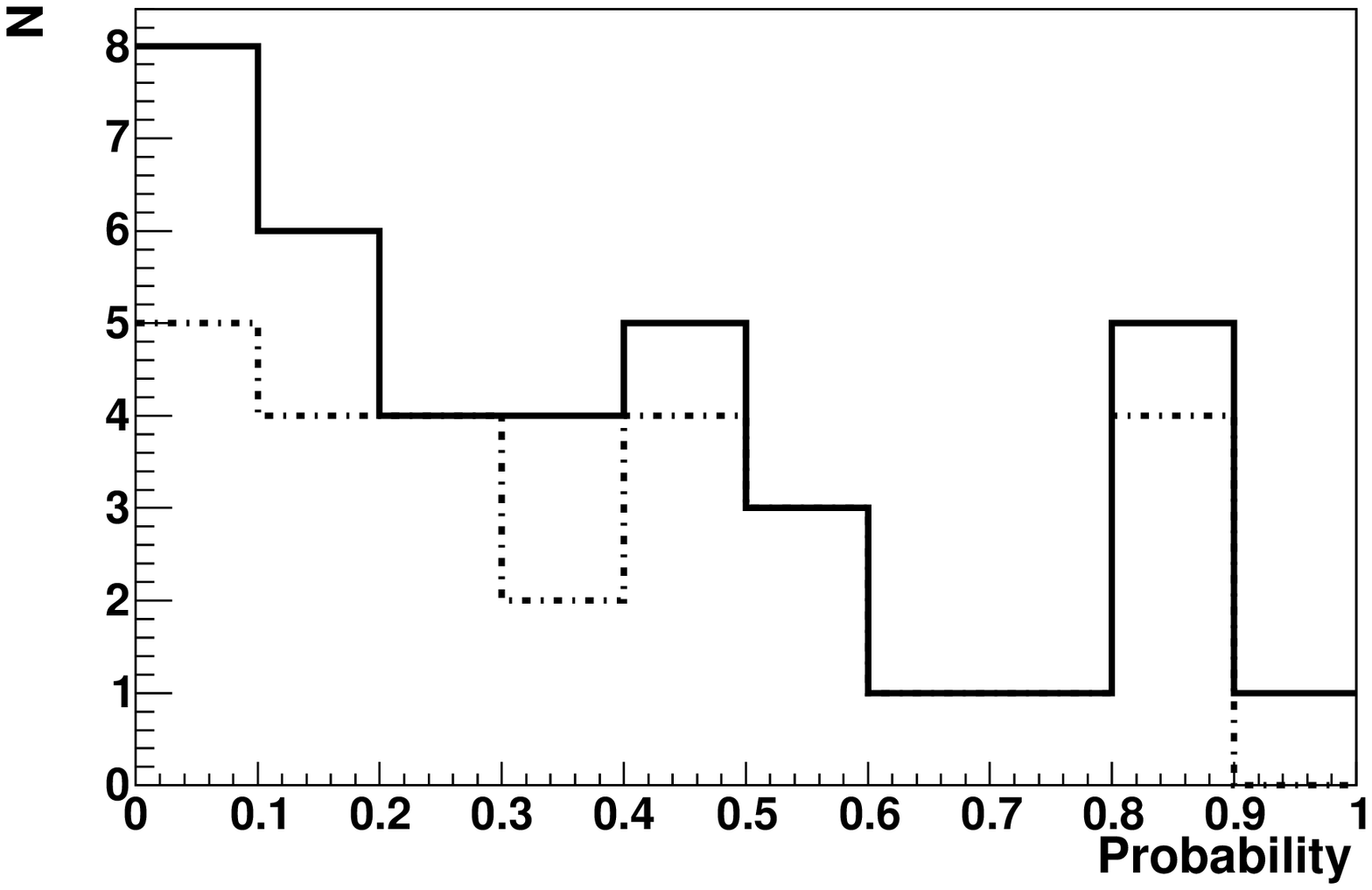} \\ [0cm]
  \mbox{\bf (a)} \\ [0cm]
  \epsfxsize=8.7cm
  \epsffile{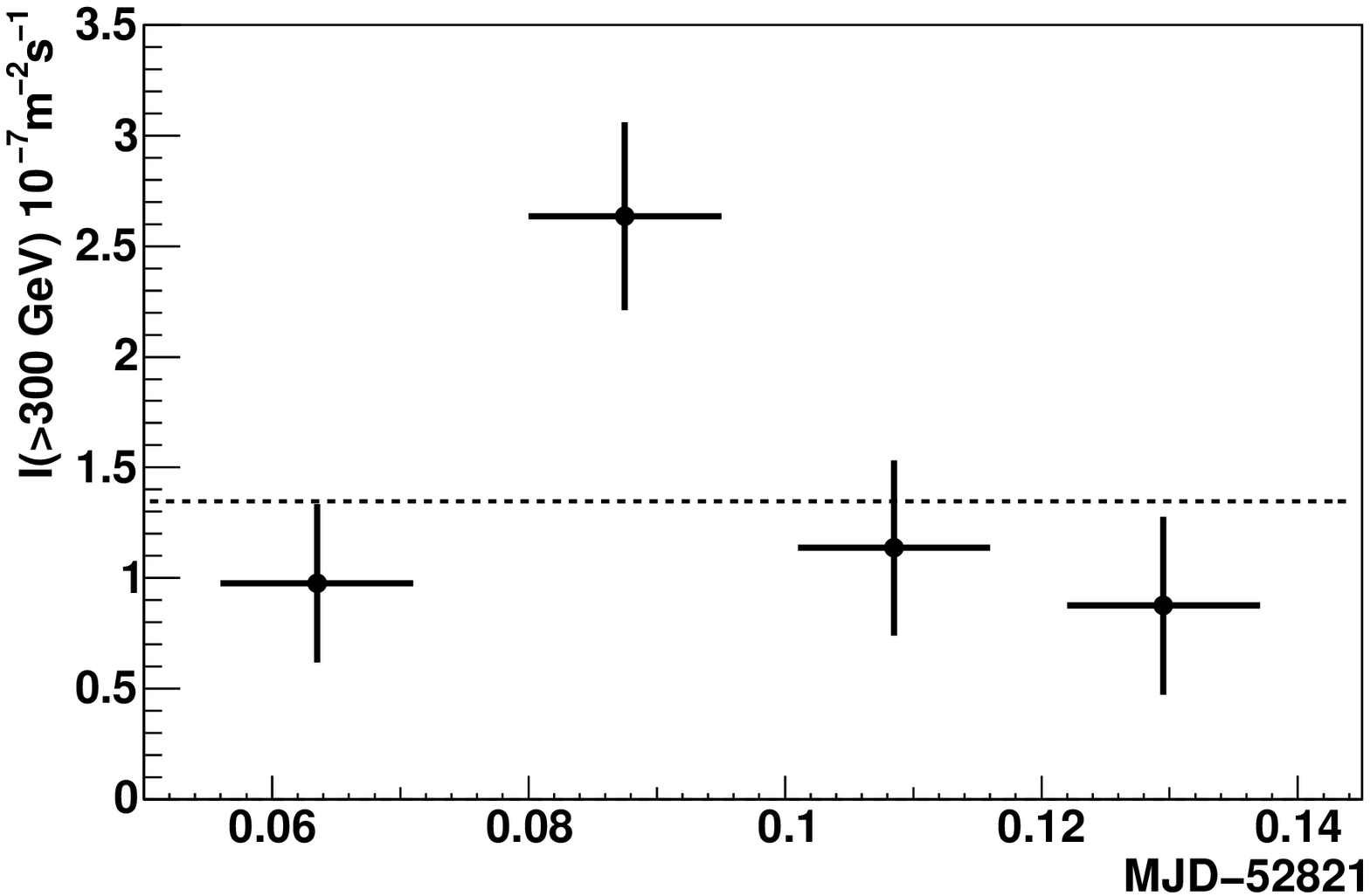} \\ [0.cm]
  \mbox{\bf (b)}

  \end{array}$
  \end{center}
  \caption{{\bf a)} The distribution of $\chi^2$ probabilities from best fits to a constant
	of the run by run integral flux greater than 300 GeV for each night
	of H.E.S.S. observations of PKS2155$-$304 in 2003. The solid line is for all nights of
	data taking and the dashed line represents nights with
	more than one degree of freedom (3 runs minimum).
	{\bf b)} The run by run integral flux greater than 300 GeV 
	observed from PKS 2155$-$304 for the night of MJD 52821.
	Only the statistical errors are shown.}	
  \label{runwisevstime}
  \end{figure}
 
A search for evidence of intra-night variability in the VHE flux from PKS2155$-$304 is performed for the 2003 data set. 
A run by run (28 min) time scale is chosen for simplicity and because each stereo run 
should yield on average $\sim$4$\sigma$ 
given that $\sim$44$\sigma$ was observed in 55.4 live time hours of stereo observation.  The study is not performed
for the 2002 data because the mono-configuration is considerably less sensitive and the data are more sparse (one on-source run
on most nights).  The respective fluxes greater than 300 GeV for each run are fit to a constant for each 
night of observation, with the hypothesis
that any variability will manifest itself in the form of a poor $\chi^2$.  Only runs with
zenith angles of less than 45 degrees are analyzed to eliminate possible systematic effects from large zenith angle
observations.  As motivated in the previous section, the assumed spectral index for the
flux calculation is $\Gamma=3.32$.  Since there are differing numbers of runs in a given night, the
$\chi^2$ probability is calculated from the fit to a constant for each night and is shown
by the solid line of Figure~\ref{runwisevstime}(a) for all nights with a minimum
of one degree of freedom.  The dashed line in Figure~\ref{runwisevstime}(a) represents the nights with more than one degree of freedom.
A larger population of nights with low probabilities exists and is evidence for intra-night flux variability.  

Figure~\ref{runwisevstime}(b) shows the flux light curve for the night with the lowest probability of being consistent with constant
along with the corresponding fit (dashed line).  On this night, MJD 52821, flaring behavior is suggested since 
the flux increases by a factor of 2.7$\pm$0.7 on the time scale of half an hour, and then decreases 
by a factor of 2.3$\pm$0.9 in a similar time.  Unfortunately at this stage in the
construction of H.E.S.S., observations were only possible 
with the less sensitive two-telescope software stereo configuration, 
thus a more detailed study is not possible due to the relatively small number of observed gamma rays ($\sim$1 per min).
However, with the completed H.E.S.S. experiment it will be possible to explore similar phenomena on a time scale of a few minutes.

\section{Conclusions}
Gamma rays with energies greater than 160 GeV have been detected with high significance ($\sim$45$\sigma$) from
the HBL object PKS 2155$-$304 in H.E.S.S. observations made in 2002 and 2003.  Although observations were
made in less sensitive configurations (fewer telescopes) of H.E.S.S. before the completion of the 
construction of the full experiment, 
a strong detection is found in each of the dark periods of observations with appreciable exposure.
The VHE flux observed from PKS 2155$-$304 is shown to be variable on the
time scale of months, nights, and hours with an average flux above 300 GeV between 10\% and 60\% of that
observed from the Crab Nebula.  
An extreme case of variability shows an increase in the observed flux by a factor of 2.7$\pm$0.7
in a 30 minute interval followed by a decrease by a factor of 2.3$\pm$0.9 in the following 30 minutes.
Energy spectra are made for each dark period for which a detection 
occurred and are found to be characterized by a steep power law. 
No evidence is found for variability in the spectral index of PKS 2155$-$304 over time, or for a
hardening of the spectrum with increased flux levels. However, these hypotheses are also not
ruled out.  A time-averaged energy spectrum is 
determined for the 2003 H.E.S.S. observations and fits to a power law with 
an exponential cutoff ($\Gamma$=$2.90^{+0.21}_{-0.23}$, 
$E_{\mathrm{cut}}$=$1.4^{+0.8}_{-0.7}$ TeV), and a broken
power law ($\Gamma_{1}$=$3.15^{+0.10}_{-0.12}$, $\Gamma_{2}$=$3.79^{+0.46}_{-0.27}$, 
$E_{\mathrm{break}}$ = $0.7\pm0.2$ TeV), show an improved 
$\chi^2$ per degree of freedom 
relative to the one obtained from a fit to a simple power-law ($\Gamma=3.32\pm0.05$).  
Although the fits to power laws with features
are not significantly better than that of a fit to a simple power law, the suggestion of features may indicate the absorption of TeV photons
on the extragalactic infrared background light.  However, any spectral curvature could also be intrinsic to the blazar.  

Comparison of the properties of PKS 2155$-$304 measured by H.E.S.S.
to those of other blazars detected at TeV energies is somewhat difficult due to the highly variable
nature of these objects and their differing redshifts. 
Regardless, comparison of the flux (magnitude and variability) and spectra, may yield 
some insights into phenomena associated with TeV bright AGN. 
In the cases of most well studied TeV blazars, Mkn 421 ($z=0.030$) and Mkn 501 ($z=0.033$), 
the observed flux has been found to vary from a fraction of the Crab flux to peak 
values of $\sim$7 ~\cite{HEGRA_421} and $\sim$10 ~\cite{HEGRA_501a} times the Crab flux respectively.  
The flux detected from 1ES 1959$+$650 ($z=0.047$) 
ranges from 5\% to 220\% of the Crab ~\cite{HEGRA_1959}, and the flux measured from  
1ES 1426+428 has shown variability by a factor of 2.5, 
with peak fluxes of order 10\% of the Crab ~\cite{HEGRA_1426b}.
Clearly, the flux observed from PKS 2155$-$304, which ranges from 10\% to 60\% of the 
Crab, does not vary as much as that from Mkn 421, Mkn 501 and 1ES 1959+650.  Additionally, the 
peak fluxes observed from these objects are considerably higher than that measured for
PKS 2155$-$304.  The observed flux from PKS 2155$-$304 is most similar, 
both in magnitude and variability, to that found from 1ES 1426$+$428.  
Comparing the spectra of the other AGN to that found from 
PKS 2155$-$304 also shows some differences.  The spectral indices, 
prior to the exponential cutoffs found, of Mkn 501 ($\Gamma=1.92\pm0.20$), Mkn 421 ($\Gamma=2.19\pm0.04$), 
and 1ES 1959$+$650 ($\Gamma=1.83\pm0.17$) are harder than $\Gamma=3.32\pm0.05$ found for PKS 2155$-$304.
Again the most similar blazar is 1ES 1426$+$428, whose spectral 
index is $\Gamma=3.50\pm0.35$ ~\cite{whipple_1426}.
It is interesting that the properties of PKS 2155$-$304 ($z=0.117$) 
are most like those of 1ES 1426$+$428, whose redshift ($z=0.129$) is also the most similar.  
However, it cannot be stated from the evidence presented that
the similarities are a direct consequence of the comparable redshifts.

As of 2004, the H.E.S.S. detector is complete and is 
considerably more sensitive than any of the configurations presented here, allowing
for more detailed studies of PKS 2155$-$304 and other astrophysical objects to be performed in the future.  

\begin{acknowledgements}

The support of the Namibian authorities and of the University of Namibia
in facilitating the construction and operation of H.E.S.S. is gratefully
acknowledged, as is the support by the German Ministry for Education and
Research (BMBF), the Max-Planck-Society, the French Ministry for Research,
the CNRS-IN2P3 and the Astroparticle Interdisciplinary Programme of the
CNRS, the U.K. Particle Physics and Astronomy Research Council (PPARC),
the IPNP of the Charles University, the South African Department of
Science and Technology and National Research Foundation, and by the
University of Namibia. We appreciate the excellent work of the technical
support staff in Berlin, Durham, Hamburg, Heidelberg, Palaiseau, Paris,
Saclay, and in Namibia in the construction and operation of the
equipment.

\end{acknowledgements}

\end{document}